\documentclass[sigconf, nonacm]{acmart}
\usepackage{bm}
\usepackage{amsmath}
\usepackage{graphicx}
\usepackage{textcomp}
\usepackage{diagbox}
\usepackage{booktabs}
\usepackage{colortbl}
\usepackage{multirow}
\usepackage{tabularx}
\usepackage{tabularray}
\usepackage{booktabs}
\usepackage{siunitx}
\usepackage{soul}
\usepackage{wrapfig}
\usepackage[caption=false]{subfig}
\usepackage[font=footnotesize,labelfont=bf]{caption}
\usepackage[normalem]{ulem}
\usepackage{tikz}
\usepackage{xcolor}
\usepackage[dvipsnames]{xcolor}
\usepackage{enumitem}

\newcommand{\circled}[2][blue!30]{%
  \tikz[baseline=(char.base)]{
    \node[
      shape=circle,
      draw=black,
      fill=#1,
      inner sep=1pt
    ] (char) {#2};
  }%
}

\newcommand{\circledblack}[2][black]{%
  \circled[#1]{\textcolor{white}{#2}}%
}
\newcommand{\EmptyCircle}{%
  \tikz[baseline=-0.5ex]\draw[line width=0.6pt] (0,0) circle (1ex);%
}
\newcommand{\FilledCircle}{%
  \tikz[baseline=-0.5ex]\fill (0,0) circle (1ex);%
}

\newcommand{\HalfFilledCircle}{%
    \tikz[baseline=-0.5ex] {
    \draw[line width=0.6pt] (0,0) circle (1ex);%
    \fill(1ex,0) arc[start angle=0, end angle=180, radius=1ex] --cycle;}
    }

\newcommand{\HalfFilledCircleLeft}{%
    \tikz[baseline=-0.5ex] {
    \draw[line width=0.6pt] (0,0) circle (1ex);%
    \fill(0,1ex) arc[start angle=90, end angle=270, radius=1ex] --cycle;}
    }
\newcommand{\NumberCircle}[2][black]{%
  \tikz[baseline=-0.5ex]
    \node[draw, circle, fill=#1, inner sep=0pt, minimum size=2.0ex, line width =0.4pt, text=black] {#2};%
}

\newcommand{\ProgressBar}[3]{%
  \tikz[baseline=-0.6ex]{
    \draw[draw=black, line width=0.4pt] (0,0) rectangle (#2,#3);
    \fill[black] (0,0) rectangle ({#1*#2},#3);
  }%
}

\let\oldcite\cite
\renewcommand{\cite}[1]{{\color{blue}\oldcite{#1}}}
\newcommand{\XRPRISM}{\textit{XR-PRISM }}

\AtBeginDocument{%
  }

\copyrightyear{2025}
\acmYear{2025}
\setcopyright{cc}
\setcctype{by}
\acmConference[VRST '25]{31st ACM Symposium on Virtual Reality Software and Technology}{November 12--14, 2025}{Montreal, QC, Canada}
\acmBooktitle{31st ACM Symposium on Virtual Reality Software and Technology (VRST '25), November 12--14, 2025, Montreal, QC, Canada}
\acmDOI{10.1145/3756884.3766045}
\acmISBN{979-8-4007-2118-2/2025/11}

\begin{document}

\title{Beyond the Headset: A Systematization of Knowledge on Extended Reality Privacy and Security in Healthcare}


\author{Nafisa Anjum}
\email{nanjum1@students.kennesaw.edu}
\affiliation{%
  \institution{Kennesaw State University}
  \state{Georgia}
  \country{USA}
}

\author{M. Rasel Mahmud}
\email{m.raselmahmud1@gmail.com}
\email{mmahmud2@kennesaw.edu}
\affiliation{%
  \institution{Kennesaw State University}
  \state{Georgia}
  \country{USA}
}

\begin{abstract}
Extended reality (XR) systems offer transformative potential for healthcare in domains ranging from surgical planning to remote rehabilitation and mental‐health therapy. The rich streams of sensor, biometric, and environmental data that enable these applications, however, also create novel and poorly understood privacy and security vulnerabilities: adversaries can exploit unencrypted signaling, sensor side‐channels, and application‐layer flaws to infer sensitive patient information or disrupt clinical workflows. Nevertheless, there aren't many thorough Systematization of Knowledge (SoK) that examine XR for healthcare at the moment. In this SoK, we survey 65 peer‐reviewed works published between 2017 and 2024 across leading XR, security, and privacy venues, synthesizing a unified threat taxonomy that spans device, network, user and cloud layers. We introduce a quantitative evaluation framework \XRPRISM (Privacy and Risk Impact Scoring Metric), drawing on adapted risk scores, detection performance, and usability assessments to rigorously assess the level of security and privacy risks. Our analysis reveals critical gaps: over 70\% of countermeasures lack standardized risk evaluations, fewer than 15\% include high prerequisites to launch an attack, and reproducibility is hampered by scarce artifact releases. Finally, we chart a research roadmap advocating for open benchmark suites with shared datasets, artifact disclosure policies, cloud‐layer protections, and robust detection and recovery mechanisms. By quantifying “what works—and by how much,” this SoK provides a data‐driven foundation for developing secure, privacy‐preserving, and usable XR healthcare technologies.
\end{abstract}




\maketitle

\vspace{-0.1in}
\section{Introduction}
Extended Reality (XR)—encompassing Virtual Reality (VR), Augmented Reality (AR), and Mixed Reality (MR), is revolutionizing healthcare by enabling immersive training, remote consultation, patient rehabilitation, and mental‐health therapies. Clinical studies have shown that VR‐based exposure therapy can reduce anxiety disorders, AR overlays can guide surgeons with sub‐millimetric precision, and mixed‐reality simulators improve motor‐skill learning in stroke rehabilitation \cite{andrews2019extended}. XR technology is revolutionizing medicine through access to care, precision treatments, and cutting-edge training. Numerous VR-based
assistive feedback improved balance and gait impairments \cite{9995441, 9756779,mahmud2022vibrotactile,mahmud2023multimodal,mahmud2022standing, mahmud2023eyes,mahmud2023visual, mahmud2023auditory, mahmud2024multimodal,cordova2023real}. However, the very data that make these applications powerful high‐fidelity motion traces, physiological signals, biometric identifiers, and rich environmental context also expose patients and providers to unprecedented security and privacy (S\&P) risks.

A few studies have been conducted so far to examine S\&P issues and offer innovative solutions to improve the security and privacy of XR experiences \cite{alhakamy2024extended,abraham2022implications,giaretta2024security}. Nevertheless, other studies have identified a disconnect between the scholarly literature on this subject and what is being done in real-world VR use cases, such as in hospitals or industry \cite{sivelle2024extended}. We contend that in order to comprehend how S\&P challenges are viewed in the XR ecosystem, what strategies are in place to handle them, and what will be required going forward to lessen the pertinent issues, a practice-based method is required.

While a growing body of work has documented individual vulnerabilities \cite{zhang2023facereader,andrade2020discerning,grichi2024biosensor}, there is no unifying framework that systematically characterizes these threats in the specific context of XR in healthcare. Our review of attack vectors (summarized in Table \ref{tab:attack-taxonomy}) reveals that most exploits require only minimal prerequisites and modest attacker expertise, making real‐world compromise alarmingly feasible.  Conversely, the defenses cataloged in Table \ref{tab:defense-approaches} are both sparse and unevenly evaluated: few offer formal guarantees, most incur high usability or performance overhead, and virtually none address detection or recovery once prevention fails.

In this Systematization of Knowledge (SoK), we aim to fill these gaps by providing a comprehensive survey and analysis of privacy and security in XR healthcare environments. Our contributions are threefold:

\begin{enumerate}[label=\circledblack{\arabic*}]
  \item We deconstruct XR into a four-layer system and provide a \textit{data-driven classification} of the numerous works, breaking them down into essential components and subcategories.
  \item Based on our threat model, we rigorously analyze a set of relevant work and map existing threats and defenses to critically evaluate their effectiveness and practical limitations.
  \item Finally, we introduce a quantitative framework, \XRPRISM that is specifically formulated to assess these security and privacy risks in terms of XR in healthcare.
\end{enumerate}
The rest of this paper is organized as follows: Section \ref{sec:background} describes the background for our four-layer systematization and Section \ref{sec:method} presents the research questions and methodology that directed our findings. Section \ref{sec:taxonomy} and Section \ref{sec:defense criterion} present our knowledge representation mechanism whereas Section \ref{sec:overview} discusses our collection of relevant works. Outlining the research gaps, we present future research directions in Section \ref{sec:challenges} along with related work in Section \ref{sec:related}. Finally, Section \ref{sec:conclusion} concludes with a call to action for the XR‐healthcare research community.

\section{Background}\label{sec:background}
XR defines the next dimension of experience in technology. The main players AR, VR, and MR\textemdash each offers a different level of immersion using cameras, sensors, and displays to change how we perceive reality. VR replaces your world, AR adds to it and MR blends with it. VR headsets transport you else where , AR overlays digital info on your phone's camera and MR makes digital objects your real space\cite{de2019security}. From gaming to professional training, each of these are built for different experiences. So, XR isn't one single thing; it's a family of tech making our interactions with computers more natural and immersive. Instead of tapping screens, we use hands, voices and movement. 
\paragraph{\textbf{Architecture}}In order to bring the XR system workflow under a common ground, we divide the XR pipeline into 4 concentric layers\textemdash User, Device, Network, and Cloud, with their key functions and data flows broken down as follows:
\paragraph{\circled{1}\textbf{User Layer}}The essential element of this layer is the human body, which is the most crucial element for interaction and the source of all intent and bio-signals.  This layer's data sources include vocal instructions, gestures, hand/controller motions, and head movements. In recent years, context-based inputs like vocal instructions, "Show me the next step," and simple button clicks, as well as biometric signals like eye gaze and EMG, have become more and more popular as explicit data\cite{han2022comic}. Another essential element, wearable/handheld interfaces, uses optical trackers and IMU sensors to record translational and rotational data as the user moves.  Pupil location is recorded using physiological data such as eye-trackers at 120–250 Hz and EMG bands sampled at 500–1 kHz \cite{rudzki2022xr, lim2024impact}.  After that, time-stamped raw data streams are queued for on-device fusion.
\paragraph {\circled{2} \textbf{Device Layer}} This layer centers on the headset and its onboard processing. It comprises embedded compute units; GPUs (Graphics Processing Unit) and DSPs (Digital Signal Processor) for audio and haptics, low-level firmware and drivers (managing sensors, displays, and peripherals), and hardware security modules (secure boot, signed firmware). First, raw IMU, optical tracker, and depth-camera data are fused—via SLAM or Kalman-filter pipelines—into a precise 6-DoF pose \cite{sheng2024review}. That pose then feeds into the local application runtime (e.g., Unity XR or OpenXR), which handles gesture recognition, UI state transitions, and any safety enforcements. The combined pose and UI events drive the scene graph update, producing the next frame. Finally, once rendering is finished, any virtual collisions or cues are converted into haptic commands and dispatched to the device’s actuators.
\paragraph{\circled{3} \textbf{Network Layer}} The key responsibility of this layer is to move data back and forth between the headset and off-device servers in a safe and quick path. First, the headset and server \textit{handshake} securely (using methods like mTLS) so they trust each other \cite{shi2024impact}. Then the headset sends streams of head and hand positions, eye-tracking, and any health signals over encrypted links (e.g., WebRTC or secure WebSockets)\cite{park2022instantxr}. Important data like positions are sent first, with video or logs coming next. Often this traffic goes through a nearby edge server to keep delays very low—useful for things like sharing a virtual room with others or running quick AI checks. The server can then send updates, analytics, or shared-session changes back to the headset on the same secure channel. To keep things smooth, the system uses buffering and error-correction, and it keeps XR traffic on its own network slice so it can’t be tampered with and private data stays protected \cite{huang2025model}.

\paragraph{\circled{4} \textbf{Cloud Layer}} Cloud layer is where heavy lifting happens off the device. First, the headset or edge server sends anonymized session data—like movement streams, eye-gaze logs, or biometric readings to the cloud over that same secure link. In the cloud, powerful AI models run analyses, for example, detecting unusual gait patterns or stress indicators and store long-term records in encrypted databases that follow healthcare policies such as HIPAA (Health Insurance Portability and Accountability Act \cite{hhs_hipaa_1996}) and GDPR (General Data Protection Regulation\cite{eu_gdpr}). The cloud can also coordinate federated learning, combining insights from many users without moving raw health data around, and push signed firmware or app updates back to devices. All cloud services sit behind strict access controls, audit logs, and micro-segmentation to keep data private and ensure only authorized systems can read or modify it\cite{park2022instantxr}.
\paragraph{\textbf{XR in Healthcare}}
Extended Reality (XR) technologies are proving transformative across a spectrum of healthcare applications, from motor rehabilitation to tele-therapy and autism interventions. In physical rehabilitation, HoloLens-based AR systems can superimpose customized 3D movement trajectories onto a patient’s workspace, yielding statistically significant improvements in kinematic precision and have high usability ratings from clinicians\cite{yang2021utilization}. Home-based tele-rehabilitation using non-immersive VR platforms has likewise shown to enhance standing function, gait, and upper-limb motor control post-stroke—randomized feasibility trials report greater improvements than app-only interventions and underscore VR’s potential for scalable, remote therapy\cite{lamb2025effectiveness}. In neurodevelopmental care, systematic reviews of XR and telehealth approaches for autism spectrum disorder (ASD) document positive outcomes in social engagement, anxiety reduction, and treatment adherence among children and adolescents\cite{chen2022extended, worlikar2023mixed}. Moreover, mobile XR applications that employ sensory-based and mediated interaction designs have demonstrated technical feasibility and preliminary effectiveness in supporting ASD therapy, suggesting a promising avenue for personalized, technology-driven care. The sensors on board XR systems provide a lot of data, which can be used to deduce other characteristics about users, such as personal and health-related characteristics like physical fitness.  In therapeutic settings, this presents serious hazards because privacy attacks could result in violation of user confidentiality \cite{mattei2017privacy}.
\section{Research Questions and Methodology}\label{sec:method}
\begin{figure}[t]
  \centering
  \includegraphics[width=0.95\columnwidth]{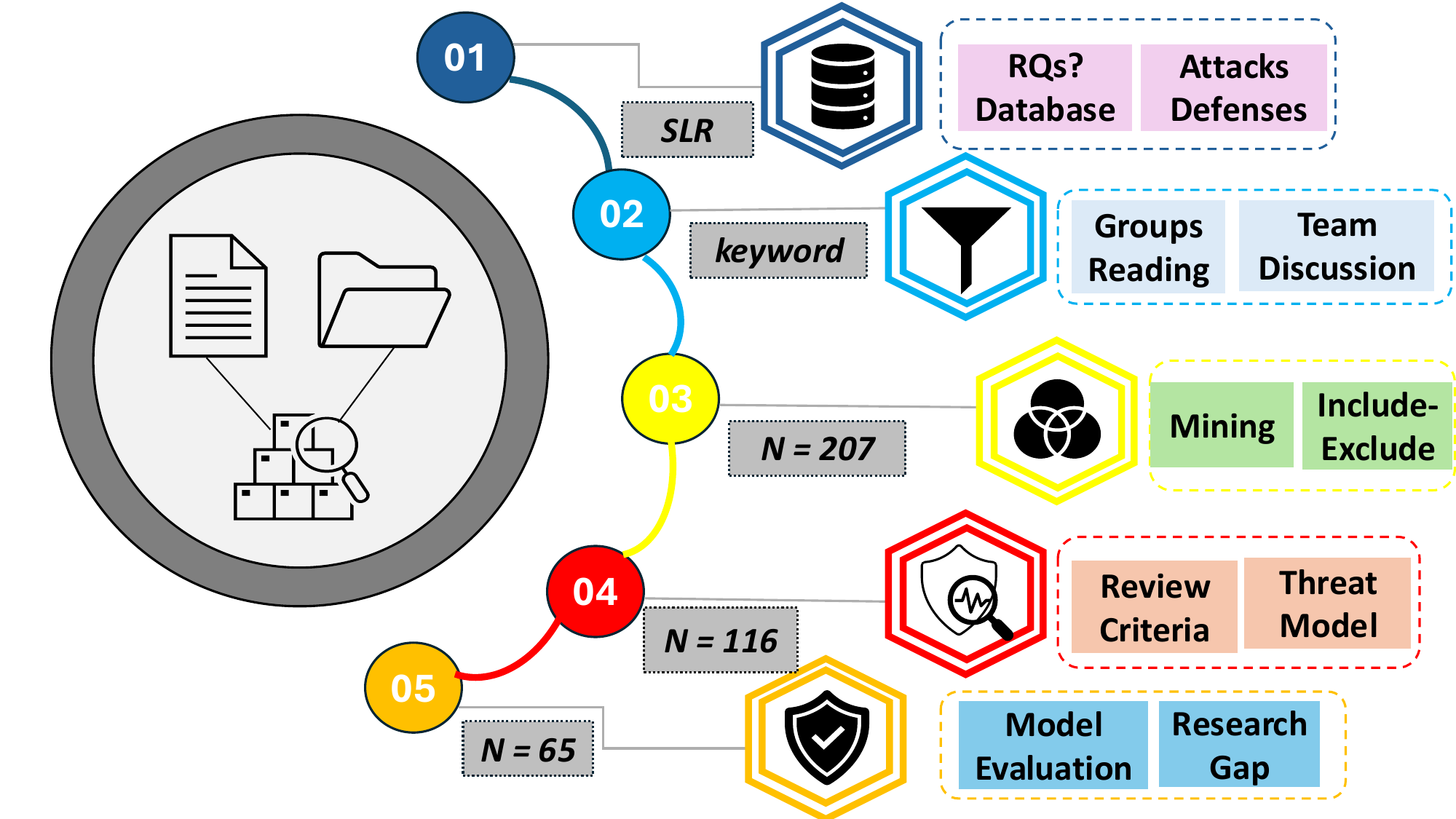}
  \caption{Methodology for Systematization of Knowledge (SoK)}
  \label{fig:survey-method}
\vspace{-0.1in}
\end{figure}

To ensure a thorough, unbiased survey of security and privacy in extended reality for healthcare, we adopt a structured, systematic literature review (SLR) process, augmented by quantitative analysis. Conducting in-depth research on the current state of security and privacy in the XR healthcare infrastructure is the goal of the Systematic Literature Review (SLR). The literature has put out a number of techniques for conducting SLRs \cite{bell2022business,creswell2016qualitative}.  We adhere to the SLR method's procedures as outlined in \cite{fink2019conducting,okoli2015guide}.  Because it offers comprehensive instructions for reviewing both quantitative and qualitative works, this approach was selected as highly suitable for this investigation \cite{ofte2023understanding}.  Additionally, as many SLRs in the literature do, the SLR complies with the Preferred Reporting Items for Systematic Reviews and Meta-Analyses (PRISMA) statement's recommendations \cite{fink2019conducting}. The following steps describe in details on how the SoK has been formulated as per Fig.\ref{fig:survey-method}.
\vspace{-0.1in}
\subsection{Research Questions}
Finding and identifying publications on various facets of S\&P in the XR healthcare domain is the aim of this literature review. In order to make recommendations for further research, this review attempts to give a thorough analysis of the area in terms of publications, studies conducted, methodologies, and emerging patterns. As a result, the following research inquiries were established:
\begin{enumerate} [label= \textbf{RQ: \arabic*.}]        
\item \textbf{What threat techniques/attacks have researchers \hspace{0pt}demonstrated against XR systems that can be applied to a healthcare setting?} \label{rq:1}
    \begin{enumerate}[label=\arabic{enumi}.\arabic*.]  
      \item Which precise system elements (e.g., IMU sensors, rendering engine, network link, haptic actuator) do these attacks focus on?\label{rq:1.1}
      \item What level of access and technical effort must an attacker muster to succeed?\label{rq:1.2}
      \item Upon successful exploitation, what are the measurable consequences for patient safety, data confidentiality or integrity?\label{rq:1.3}
    \end{enumerate}
  \item \textbf{What defensive strategies have been proposed to secure XR environments?}\label{rq:2}
    \begin{enumerate}[label=\arabic{enumi}.\arabic*.]
      \item  Which parts of the XR-health pipeline (from device firmware to UI consent dialogs) have existing countermeasures been designed to protect?\label{rq:2.1}
      \item What operational or performance costs (such as hardware upgrades, added latency) are associated with deploying these defenses?\label{rq:2.2}
    \end{enumerate}
    \item \textbf{To what extent do the available defenses align with the catalog of known attacks—where are the gaps in coverage, and which threat scenarios remain unaddressed?}\label{rq:3} 
\end{enumerate}
\subsection{Methods for Systematization}
\paragraph{\circled{1} \textbf{Academic Databases}}
The \textit{IEEE Xplore}, \textit{ACM Digital Library}, \textit{USENIX Proceedings} and \textit{PubMed} are the academic research databases where the SLR is conducted.  All IEEE-VR, ISMAR, VRST proceedings and journals which are the most popular for XR research are indexed in IEEE Xplore. Besides, ACM contains CHI, UIST, VRST papers and USENIX covers top-notch security papers. PubMed has been utilized for clinical XR applications, medical device security and regulatory compliance. Based on the guidelines for conducting literature reviews, these databases have been chosen as the most suitable for this investigation.  
\paragraph{\circled{2} \textbf{Keyword Mapping}}
\begin{table}[t]
  \footnotesize
  \setlength{\tabcolsep}{4pt}
  \caption{Keyword Strategies for XR-Healthcare}
  \label{tab:keyword-strategies}
  \begin{tabular}{@{}l p{0.68\columnwidth}@{}}
    \toprule
    \textbf{Groups}      & \textbf{Terms} \\
    \midrule
    XR Modality            & virtual reality \textbf{OR} augmented reality \textbf{OR} mixed reality \textbf{OR} extended reality \\
    Healthcare Context     & healthcare \textbf{OR} medical \textbf{OR} tele-rehabilitation \textbf{OR} surgical \textbf{OR} therapy \\
    Security Focus         & security \textbf{OR} attack \textbf{OR} threat \textbf{OR} vulnerability \textbf{OR} side-channel \\
    Privacy Focus          & privacy \textbf{OR} data protection \textbf{OR} differential privacy \textbf{OR} PHI \\
    Defense Mechanisms     & mitigation \textbf{OR} countermeasure \textbf{OR} access control \textbf{OR} encryption \textbf{OR} obfuscation \\
    Quantitative Metrics   & risk assessment \textbf{OR} CVSS \textbf{OR} DREAD \textbf{OR} differential privacy \textbf{OR} latency \textbf{OR} usability \\
    \bottomrule
  \end{tabular}
\vspace{-0.1in}
\end{table}
In order to select search terms, at first 6 groups were formed: \circledblack{1} XR Modality, \circledblack{2} Healthcare Context, \circledblack{3} Security Focus, \circledblack{4} Privacy Focus, \circledblack{5} Defense Mechanisms, \circledblack{6} Quantitative Metrics; the details of the terms in each group are included in Table {\ref{tab:keyword-strategies}}. Based on the combination of these keywords, we used search queries such as: virtual reality \textbf{AND} healthcare \textbf{OR} PHI (Protected Health Information) \textbf{AND} security, XR \textbf{AND} rehabilitation \textbf{AND} side-channel \textbf{AND} differential privacy, extended reality \textbf{AND} CVSS (Common Vulnerability Scoring System\cite{first2019cvss}) \textbf{OR} DREAD (Damage, Reproducibility, Exploitability, Affected users, Discoverability\cite{wikidread}).
\paragraph{\circled{3} \textbf{Screening}}
The majority of the academic work on XR privacy and security was released after 2017 \cite{patel2022systematic}. Furthermore, the HTC Vive line of HMDs (Head-Mounted Display) and the Meta Quest (previously Oculus), two of the most well-known VR devices available today, were not made available to the general public until 2016 \cite{wikipedia_htcvive_2025}. In order to maintain the relevance of the results, 2017 was selected as the threshold year and any studies published before that year were not included. The search returned 207 records spanning from 2017 to 2024. We removed 91 duplicates and irrelevant studies, yielding 116 unique records. Subsequently, inclusion and exclusion principles were applied.
\paragraph{\circled{4} \textbf{Assessment for final scope}}
To ensure our survey covers only those XR security and privacy works most relevant to healthcare applications, we applied the following eligibility rules:

\textbf{Inclusion Criteria.}
Studies that explicitly analyze threats or propose defenses—attacks, side-channels, authentication, data protection, anomaly detection, or recovery mechanisms—in XR environments that can also be applied to the healthcare industry were considered as the prime focus.This is because not many studies show security experiments conducted in a clinical testbed which we have discussed in Sec.\ref{sec:challenges}. We conducted a thorough qualitative synthesis of the 116 papers that were chosen, concentrating on their attack or defense methodologies, the impact, overhead and accuracy of attack or defense reported.  Every paper was carefully examined to obtain these information by observing the mathematical formulations, and visual aids such as tabular representations and figures or diagrams that highlight significant contributions. Following analysis, we removed 51 papers since they did not meet our criteria for additional study analysis. Our GitHub \footnote{\url{https://github.com/User32-blip/SoK-XR-in-Healthcare}} now provides a list of the 65 papers that were chosen for the study.

\textbf{Exclusion Criteria.}
General XR security/privacy papers without discussion or implications for healthcare or medical data such as security for automotive applications, learning platforms or network security \cite{sofiadin2024threats,pappas2021extended} were not included in this study. Papers on device design, tracking accuracy improvements, or system throughput that do not address threat models or privacy risks along with non peer-reviewed or incomplete artifacts, theses, white papers, or demos lacking sufficient methodological detail were also excluded. Additionally, publications in languages other than English, or dated prior to January 2017 were not included in the scope of this SoK.

\begin{figure}[t]
  \centering
  \includegraphics[width=\columnwidth]{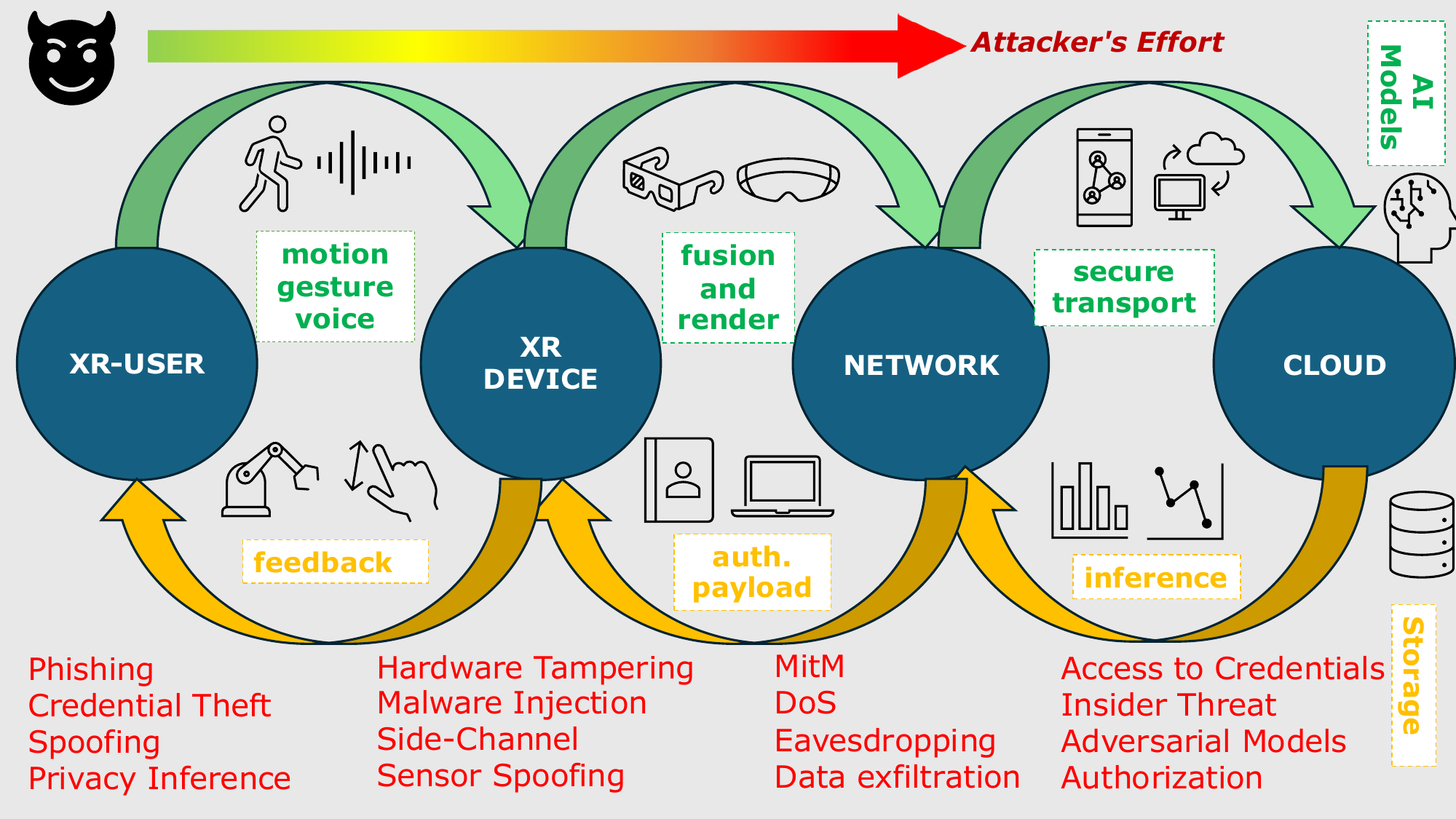}
  \caption{Threat model of an XR system.}
  \label{fig:xr-threat-model}
  \vspace{-0.2in}
\end{figure}

\paragraph{\circled{5} \textbf{Data Analysis for Modeling}}
We first conducted a thorough review of all selected XR security and privacy publications to extract the core attributes of each work—namely, the XR modality and system component under threat (device, network, user, or cloud layer), the specific attack vectors and adversary capabilities (prerequisites and required expertise), and the characteristics of each proposed defense (preventative, detective, or recovery group; mitigation vector; overhead; maintenance cost; and impact). The outcome of this data analysis and modeling was the identification of the classification criteria covered in Sec.\ref{sec:taxonomy}, such as the XR target component and attack prerequisites, as well as the building blocks for the initial draft of our threat model, which is shown in Fig.\ref{fig:xr-threat-model}. 
\paragraph{\circled{6} \textbf{Model Evaluation}}
In order to generate the SoK contributions, we assess the data, threat model, and criteria created in phase 5.  These findings serve as a summary of our findings and help us pinpoint open challenges that were covered in Sec.\ref{sec:challenges}. Lastly, we assess the findings to create the quantitative framework, \XRPRISM, that is covered in Sec. \ref{sec:xr-prism}.

\section{Evaluation Criterion: Threats}\label{sec:taxonomy}
This subsection introduces the core axes along which we categorize every XR-health attack in this SoK for answering our \ref{rq:1}
\vspace{-0.1in}
\subsection{ICS-Aligned Threat Taxonomy}
Because XR-health systems tightly couple digital controls with physical patient outcomes, the MITRE ATT\&CK for ICS Matrix \cite{mitre_attack_ics} maps very naturally onto their threat surface. We treat the XR "pipeline", signals~$\bm{\to}$sensors~$\bm{\to}$network/backend~$\bm{\to}$application, much like an ICS process-control flow and then slot each XR threat into the corresponding ICS tactic. This threat model organizes "tactics" which indicate the adversary’s high-level goals or the “why” and "techniques", the specific methods or the “how” into a matrix that defenders can use to identify, classify, and mitigate threats. The 12 ICS tactics cover the full attack lifecycle—from Initial Access through Execution, Persistence, Collection, to Impact, ensuring we didn't omit important adversary goals or techniques when mapping XR-specific threats. Using this as a backbone makes it easy to spot which areas lack XR-health research or defenses—for instance, if no papers address \textit{Privilege Escalation} or \textit{Persistence} in headset firmware, that gap becomes immediately visible. Additionally, many medical-device and safety standards, including FDA, IEC 62304 Software Lifecycle Standard \cite{FDA2023CybersecurityMedicalDevices}, already reference ICS-style threat assessments. 
\vspace{-0.1in}
\subsection{Threat Dimensions}
This subsection helps understand not only \textit{what} the threat is, but \textit{where} it strikes, \textit{how} it operates, and \textit{what} it targets.
\paragraph{\textbf{Architectural Layer.}}
The stage of the XR architecture that the threat exploits is one of the most important aspects for classification. Each threat was classified in this method as it clarifies whether the adversary is interfering with raw sensor streams and exploiting XR hardware and physical surroundings (\textit{Device}), tampering with network traffic (\textit{Network}) or application logic (\textit{Cloud}), manipulating what the user sees or interacts with (\textit{User}). This clarifies the layer from which the risk originates.
\paragraph{\textbf{Attack vector}}
The vector represents the mechanism or channel used to carry out the attack such as side-channel, network injection, UI spoofing, firmware manipulation. It highlights \textit{how} an adversary gains influence—whether it’s eavesdropping on IMU readings, performing a man-in-the-middle on streaming video, or planting a backdoor in device firmware. 
\paragraph{\textbf{XR Target Component}}
Each attack method was also categorized based on the XR system component that is the primary target. The authors' precise descriptions in the papers served as the basis for retrieving this information.
\vspace{-0.1in}
\subsection{Threat Complexity}
In order to systematically compare and prioritize XR-health attacks, we assess each technique along two orthogonal axes: \circledblack{1} the adversary prerequisites (including required access and authentication), and \circledblack{2} the expertise required to launch the attack. These evaluations feed directly into the CVSS “Privileges Required” and “Attack Complexity” metrics\cite{first2019cvss}, as well as the DREAD “Exploitability” score \cite{wikidread}.
\paragraph{\textbf{Adversary prerequisites}}
Every attack is characterized by the minimum access the adversary must possess. Some side-channel techniques such as motion-based fingerprinting\cite{giaretta2024security}, require the attacker to be within wireless range of the headset’s IMU transmissions or to attach a probing device to the hardware for high-fidelity capture.
Man-in-the-middle and replay attacks on 3D streams need the attacker to intercept or inject packets on the same LAN or VPN as the XR device \cite{de2019security}, but no local device compromise.
Overlay injection typically exploits valid application-level tokens or credentials to submit malicious rendering commands.
Consent-dialog spoofing requires tricking a user into approving data-sharing prompts, borrowing tactics from phishing in enterprise settings\cite{koch2023ok}.

We structured the authentication requirements to align with our prerequisites analysis in Table \ref{tab:auth-requirements}. For instance, no authentication or special access required is taken as Low, if valid user credential is required, the prerequisite is Medium whereas privileged access such as admin/root rights/ biometric auth are considered High level.
\begin{table}[t]
  \footnotesize
  \setlength{\tabcolsep}{4pt}
  \caption{Authentication Requirements}
  \label{tab:auth-requirements}
  \begin{tabular}{@{}l p{0.75\columnwidth}@{}}
    \toprule
    \textbf{Level}       & \textbf{Description} \\
    \midrule
    None                 & No credentials or user interaction required  \\
    User-Level           & Requires a valid user session token/login \\
    Admin/Root           & Needs elevated OS or firmware privileges on the XR device \\
    Biometric/MFA        & Must bypass multi-factor or biometric protections \\
    \bottomrule
  \end{tabular}
  \vspace{-0.2in}
\end{table}
Thus, the \textit{Requisite} column is represented by \NumberCircle[yellow]{1} as low-level, \NumberCircle[orange]{2} as mid-level and \NumberCircle[red]{3} as high-level prerequisites to conduct an attack.
\paragraph{\textbf{Attacker's Expertise}}
Based on the prerequisites and skills such as  multithreaded synthetic input injection, proficiency in signal preprocessing, signal reconstruction, coordinate-system alignment, saliency detection, and side-channel trace analysis that an attacker must muster to launch an attack successfully, the \textit{Expertise} is mapped as (\EmptyCircle) for low-level, (\HalfFilledCircle) for mid-level and (\FilledCircle) for high-level.
\vspace{-0.1in}
\subsection{Impact}
Motion-based fingerprinting of head-tracking streams can re-identify patients and infer medical conditions purely from sensor data \cite{xu2024review}. Contactless side-channel attacks reveal which $360^{\circ}$ video a patient is watching, leaking therapy content\cite{nguyen2024penetration}. Outward-facing cameras inadvertently record bystanders or map sensitive equipment layouts, breaching privacy and raising HIPAA concerns \cite{giaretta2024security}. A recent scoping review highlights that many clinical XR deployments lack encryption or consent-management controls, exposing providers to regulatory penalties \cite{lake2024cybersecurity}. As such, the degree to which an attack exposes sensitive patient data (e.g., head-tracking patterns as biometric identifiers, session recordings) has been included as a metric here to consider the impact which has been depicted via a progress bar.
\vspace{-0.1in}
\section{Evaluation Criterion: Defenses}\label{sec:defense criterion}
Following the explanation of our attack requirements, we aim to answer \ref{rq:2} in this section, discussing the defenses.
\paragraph{\textbf{Defense Group}}
Defenses have been mainly grouped into three categories: \circledblack{1} Data Obfuscation \circledblack{2}Access Control and \circledblack{3}\hspace{0pt}Authentication. The attacks originating from each of the risk layers might not have the same layer consisting of the defense mechanism and most mechanisms are developed for the \textit{Device} layer\cite{yang2024can,slocum2023going}. Hence, the grouping has been done based on their implementations.
\paragraph{\textbf{Mitigation}}
This describes the process through which the defense is implemented such as using noise, limiting access to IMU or designing new APIs. It offers an insight on which path to take for specific attacks and to what extent it prevents the attack from being successful.
\paragraph{\textbf{Defense Vector}}
The route taken to prevent the attack payload from being delivered is the defense vector. The vectors of attack and defense will typically coincide. But they may not be the same.  For instance, the randomized keyboard is a \textit{Device} based defense mechanism used by Wang et al\cite{wang2024gazeploit}.
\vspace{-0.1in}
\subsection{Deployability}
The following standards assess the difficulty of implementing a defense strategy to safeguard XR systems in accordance with RQ:\ref{rq:2.2}
\paragraph{\textbf{Trade-off}}The defense method's overhead such as additional hardware components, new system configuration and trade-off from implementation such as lower immersion, high latency, is measured by this criterion, and it can be either zero (\EmptyCircle), negligible (\HalfFilledCircleLeft) or significant (\FilledCircle).
\paragraph{\textbf{Maintenance}}
This criterion measures how much post deployment maintenance of a certain defense strategy is required.
A defense strategy may need continuous maintenance (\NumberCircle[red]{3}),intermittent maintenance (\NumberCircle[yellow]{2}),  or very negligible maintenance (\NumberCircle[green]{1}).
\vspace{-0.1in}
\subsection{Defense Robustness}
To understand the extent to which the defense is successful, we categorize as follows:
\paragraph{\textbf{Efficacy}}
The efficacy of the defense strategy indicates how effective it is and the papers themselves provide the accuracy values. If the accuracy was not reported or specified, most papers reported the percentage to which the attack success rate (ASR) decreased. Since accuracy is the most frequently mentioned metric in defense method studies, we chose it.  
\paragraph{\textbf{Stage}}
Three distinct phases of the security process can be used to classify defenses.\circledblack{1} The goal of Prevention Defense ($P$) is to lessen the likelihood of an incident occurring before a known vulnerability is taken advantage of, \circledblack{2} Detection Defense ($D$) aims to identify the attack while its being conducted and \circledblack{3} Recovery mechanism ($R$) seeks to lessen the harm caused by an attack once it has been executed effectively.
\section{Outline of Attacks and Defenses}\label{sec:overview}
\paragraph{\textbf{Target: Device Layer}}
In shared virtual environments, \textit{Yang et al.}\cite{yang2024can} illustrate this by creating and executing a novel class of keystroke inference attacks that allow an attacker (VR user) to retrieve material typed by another VR user by looking at their avatar. These attacks correctly identify 86\%–98\% of typed keys for 13 out of 15 tested users, and the recovered content preserves up to 98\% of the original typed content's context. In contrast to previous research \cite{balzarotti2008clearshot, banerjee2012biometric}, this approach simply needs a noisy virtual reality depiction of the target's hand movements rather than actual physical observation of the target. Since no authentication is required for access to the telemetry data, the authors suggested this preventative measure as a defense. \textit{Slocum et al.}\cite{slocum2023going} demonstrate how an attacker can easily extract stream head tracking data from an AR/VR device, segment it, and categorize it to get private text data. The attack is also resistant to less drastic mitigation techniques, like lowering the frequency and accuracy of sensor readings to the point where the background application would need to be compromised visually before effective mitigation of the attack. A non-intrusive privacy assault, \textit{FaceReader}\cite{zhang2023facereader} was designed by \textit{Zhang et al.}, that uses unrestricted AR/VR motion sensors to reconstruct high-quality vital sign data, such as breathing and heartbeat patterns. The foundation of \textit{FaceReader} is the fundamental realization that when the headset is positioned closely on the user's face, motion sensors may pick up on minute facial vibrations brought on by the user's heartbeat and breathing. The findings show that the attack can correctly identify gender (above 93.33\%) and recreate vital signs with low mean errors. Such risks pose serious privacy concerns in healthcare as it may leak PHI and provide biased treatment based on demographic inference \cite{ferrer2021bias}. Although some studies recommend lowering the sampling rate of signals to reduce the attack accuracy \cite{slocum2023going, yang2024can}, this might lower the user experience and immersive nature of XR \cite{zielinski2015exploring}. 
\paragraph{\textbf{Target: User Layer}}
When engrossed in virtual reality (VR), individuals have very little awareness of the outside world, according to survey study \cite{giaretta2024security}.  This makes it simple for someone with bad intentions to capture video of the targeted user's hand movements. \textit{Gopal et al.}\cite{gopal2023hidden} introduces Hidden Reality \textit{(HR Model)}, a video-based side-channel attack that demonstrates how, even if the virtual screen in VR devices is not directly visible to adversaries, indirect observations could be used to acquire the user's personal data. The \textit{GAZEploit}\cite{wang2024gazeploit} attack exploits the user's gaze behavior, which is a biometric and behavioral signal tied directly to the user's interaction. While the attack can infer over 80\% keystrokes remotely, it needs a gaze-controlled keystroke input method with other requirements which are currently available only in the Apple Vision Pro \cite{apple_vision_pro}. This indicates high prerequisites to conduct the attack. As a countermeasure, the virtual keyboard could be randomized but this would significantly impact the typing speed and accuracy of the users\cite{jiang2022learning}. \textit{Meng et al.} introduce \textit{AvatarHunter}\cite{meng2024anonymizing}, a novel attack that uses behavioral biometrics such as head and hand motion trajectories to uniquely identify users across different VR sessions. Introducing noise and obfuscating the motion data results in the Attack Success Rate(ASR) falling from 92\% to almost 50\%. \textit{Wang et al.}\cite{wang2021nod} developed \textit{Node to Auth}, a lightweight, fluent authentication system for AR/VR platforms that uses head and neck movement patterns as a behavioral biometric. The system captures natural motion during normal use; without requiring any explicit user interaction and performs continuous, real-time user authentication. This type of work has also been done by \textit{Steil et al.}\cite{steil2019privacy} who  implemented a privacy-preserving framework for eye-tracking data using differential privacy (DP). 
\paragraph{\textbf{Target: Network Layer}}
\textit{Arafat et al.}\cite{al2021vr} introduced \textit{VR-Spy}, a brand-new human activity-based side-channel attack that infers text inputs from virtual reality devices.  VR-Spy detects and identifies the microactivity linked to virtual keyboards by taking advantage of changes in the Channel State Information (CSI) of ubiquitous indoor WiFi signals. But the prequisites are high as the threat model assumes the attacker has access to the WiFi access point to obtain the CSI values. The \textit{Hijacking} attack\cite{yarramreddy2018forensic} against Bigscreen exploits the fact that much of its client-server communication (including WebSocket handshakes and ICE/STUN exchanges) is sent in cleartext over TCP. An attacker can sniff session identifiers in cleartext, perform a Man-In-The-Middle (MitM) to inject themselves or impersonate or manipulate session data. \textit{Rafique et al.}\cite{rafique2020tracking} discussed the \textit{Jamming} attack on the HTC Vive’s optical tracking which exploits its reliance on unprotected infrared (IR) sync pulses. The attacker emits IR “fake” sync pulses with randomized on-times (62–135 ms) and off-times (265–8000 ms).
The \textit{ARSpy}\cite{shang2020arspy} leverages the observable metadata of encrypted network traffic in location-based AR apps to pinpoint and track users without needing any location permissions. 
\paragraph{\textbf{Target: Cloud Layer}}
\textit{Tseng et al.}\cite{tseng2022dark} exploited the fact that VR platforms typically trust any code running in a VR app with direct, unmediated access to the user’s body-tracking streams and environment model. The Run Malicious Code attack in VR hinges on getting arbitrary code execution on the user’s headset by tricking them into installing or by covertly loading a malicious VR application. Once that code is running, it can invoke any of the perceptual–physical manipulation techniques to hijack the user’s bodily movements. Confidential user information (avatars, session participation, potentially even emotion-tracking data) is exposed through the in the vSocial VRLE testbed\cite{gulhane2019security} where an attacker can carry out a packet-sniffing attack simply by capturing the unencrypted traffic between clients and the VRLE server. In contexts involving protected populations (youth with learning disabilities), such data leakage can run afoul of privacy laws or institutional policies.
\begin{table*}[t]
  \centering
  \footnotesize
  \caption{An overview of XR Attack Methods (Number of studies included as (n) )}
  \label{tab:attack-taxonomy}
  \begin{tabular}{@{}lllllccl@{}}
    \toprule
    \multirow{2}{*}{\textbf{Approach}} 
      & \multirow{2}{*}{\textbf{Layer}} 
       & \multirow{2}{*}{\textbf{Technique}} 
      & \multicolumn{2}{c}{\textbf{Dimension}} 
      & \multicolumn{2}{c}{\textbf{Complexity}}  
      & \multirow{2}{*}{\textbf{Impact}} \\
    \cmidrule(lr){4-5} \cmidrule(lr){6-7}
      &  &  &  \textbf{Attack Vector} & \textbf{Component} 
      & \textbf{Requisite} & \textbf{Expertise} &   \\
    \midrule
    Keystroke Inference Attacks (1) {\cite{yang2024can}} & \multirow{20}{*}{Device}       & Keystroke Inference       & Side-Channel        & Hand Motion Telemetry  & \NumberCircle[yellow]{1}     & \HalfFilledCircle  & \ProgressBar{0.85}{3em}{0.8ex}   \\
    
    FaceReader (1){\cite{zhang2023facereader}} & &Eavesdropping &AR/VR headset IMU &Unrestricted Motion Sensors &\NumberCircle[orange]{2} &\HalfFilledCircle & \ProgressBar{0.97}{3em}{0.8ex} \\

    TyPose (1){\cite{slocum2023going}} & &Keystroke Inference &IMU/head tracking &On-device motion sensors &\NumberCircle[red]{3} &\FilledCircle &\ProgressBar{0.8}{3em}{0.8ex} \\
    User Identification (4)\cite{nair2023unique,moore2021personal,pfeuffer2019behavioural,liebers2021understanding} & &Inference from motion &IMU/VR motion telemetry &Motion sensors &\NumberCircle[orange]{2} &\HalfFilledCircle &\ProgressBar{0.9}{3em}{0.8ex}\\
    LensHack (1)\cite{khalili2024virtual} & &Keystroke Inference &Video side-channel &External camera &\NumberCircle[orange]{2} &\FilledCircle &\ProgressBar{0.83}{3em}{0.8ex}\\
    HoloLogger (1)\cite{luo2022holologger} & &Monitoring head motion &Malicious MR App &HMD &\NumberCircle[orange]{2} &\HalfFilledCircle &\ProgressBar{0.93}{3em}{0.8ex}\\
    User Profiling (4)\cite{tricomi2023you,nair2023truth,nair2023inferring,olade2020biomove} & &Inference from motion &Head motion &HMD &\NumberCircle[yellow]{1} &\HalfFilledCircle &\ProgressBar{0.85}{3em}{0.8ex}\\
    Chaperone (1)\cite{valluripally2020attack} & &VR boundaries &Spatial boundary APIs &HMD &\NumberCircle[orange]{2} &\HalfFilledCircle &\ProgressBar{0.8}{3em}{0.8ex}\\
    Human Joystick (1)\cite{casey2019immersive} & &Redirecting user paths &Motion guidance &Spatial rendering &\NumberCircle[orange]{2} &\HalfFilledCircle &\ProgressBar{0.8}{3em}{0.8ex}\\
    Snooping (1)\cite{wu2023privacy} & &Keystroke Inference &Side- Channel & Sensors &\NumberCircle[orange]{2} &\HalfFilledCircle &\ProgressBar{0.8}{3em}{0.8ex}\\
    Shared State AR (1)\cite{slocum2024doesn} & &Behavioral inference &Implicit Sensor Access &HMD \& controller &\NumberCircle[orange]{2} &\HalfFilledCircle &\ProgressBar{0.8}{3em}{0.8ex}\\
    Concurrent-App Fingerprinting (1)\cite{zhang2023s} & &Launch-induced patterns &Side-Channel &Unreal Engine RHI &\NumberCircle[orange]{2} &\FilledCircle
    &\ProgressBar{0.85}{3em}{0.8ex}\\
    Bystander Ranging (1)\cite{zhang2023s} & &Update spatial meshes &Side-Channel &Spatial-mapping &\NumberCircle[red]{3} &\FilledCircle
    &\ProgressBar{0.8}{3em}{0.8ex}\\
    Biosensor (1)\cite{grichi2024biosensor} & &EEG-based Identification &Raw EEG data &HMD &\NumberCircle[orange]{2} &\HalfFilledCircle
    &\ProgressBar{0.5}{3em}{0.8ex}\\
    Mobile AR (1)\cite{lehman2020stealthy} & &Stealthy Frame‐Capture &High‐level AR library &ARCore rendering &\NumberCircle[yellow]{1} &\HalfFilledCircle
    &\ProgressBar{0.6}{3em}{0.8ex}\\
    INTRUDE (1)\cite{nguyen2024penetration} &  & Deep-learning  &Visual side-channel &External camera &\NumberCircle[orange]{2} &\FilledCircle
    &\ProgressBar{0.95}{3em}{0.8ex}\\
    Implicit Identification (2)\cite{liebers2021using,miller2022combining} &  &Gaze–Head–Based & VR sensor APIs &IMU pipeline &\NumberCircle[yellow]{1} &\HalfFilledCircle
    &\ProgressBar{0.75}{3em}{0.8ex}\\

    VR-Gear (1)\cite{ling2019know} &  &Computer Vision based &Video side-channel &HMD touchpad &\NumberCircle[orange]{2} &\HalfFilledCircle &\ProgressBar{0.65}{3em}{0.8ex}\\
    Face-Mic (1)\cite{shi2021face} &  &Eavesdropping &Sensor data API &Motion sensors &\NumberCircle[yellow]{1} &\FilledCircle &\ProgressBar{0.93}{3em}{0.8ex}\\
    Keylogging (1)\cite{meteriz2022keylogging} &   &Key-tap inference &Hand-tracking data &Air-tap keyboard &\NumberCircle[yellow]{1} &\HalfFilledCircle &\ProgressBar{0.92}{3em}{0.8ex}\\
    
    \midrule
    Hand gesture (1)\cite{gopal2023hidden} &\multirow{6}{*}{User} &Exploit typing gesture &Side-Channel &Video segment &\NumberCircle[red]{3} &\FilledCircle &\ProgressBar{0.7}{3em}{0.8ex}\\
   
    AvatarHunter (1)\cite{meng2024anonymizing} & &Exploit avatar motion &Side-Channel &VR Avatars &\NumberCircle[yellow]{1} &\HalfFilledCircle &\ProgressBar{0.9}{3em}{0.8ex}\\

    Password theft (1)\cite{shukla2019stealing} & &Exploit hand gesture &Video side-channel &Touch-keyboard UI &\NumberCircle[orange]{2} &\FilledCircle &\ProgressBar{0.7}{3em}{0.8ex}\\
    
    GAZEploit (1)\cite{wang2024gazeploit} & &Remote Keystroke Inference &Gaze typing &Video segment &\NumberCircle[red]{3} &\HalfFilledCircle &\ProgressBar{0.8}{3em}{0.8ex}\\
    Shoulder Surfing (2)\cite{adams2018ethics,abdrabou2022understanding} & &Visual surveillance &Physical or screen-based & User interface &\NumberCircle[yellow]{1} &\HalfFilledCircle &\ProgressBar{0.7}{3em}{0.8ex}\\
    EyeTell (1)\cite{chen2018eyetell} & &Video-assisted gaze &video side-channel &Eye-tracking system &\NumberCircle[yellow]{1} &\HalfFilledCircle &\ProgressBar{0.85}{3em}{0.8ex}\\
    \midrule
     VR-Spy (2)\cite{al2021vr,fu2018writing} &\multirow{5}{*}{Network} &Wireless Sniffing &Virtual Keystrokes &CSI data &\NumberCircle[red]{3} &\FilledCircle &\ProgressBar{0.69}{3em}{0.8ex}\\
     Denial of Service (1)\cite{valluripally2021modeling} & &Overload of requests &Network session &Local app &\NumberCircle[yellow]{1} &\HalfFilledCircle &\ProgressBar{0.7}{3em}{0.8ex}\\
    Hijacking (2)\cite{yarramreddy2018forensic,trimananda2022ovrseen} & & Token theft &Session token &Session backend &\NumberCircle[orange]{2} &\FilledCircle &\ProgressBar{0.7}{3em}{0.8ex}\\
    Jamming (1)\cite{rafique2020tracking} & & Radio signal interference &Wireless spectrum &Network interface &\NumberCircle[orange]{2} &\FilledCircle &\ProgressBar{0.85}{3em}{0.8ex}\\
    ARSpy (1)\cite{shang2020arspy} & &Triangulation &Proximity updates &Network protocol &\NumberCircle[orange]{2} &\HalfFilledCircle &\ProgressBar{0.7}{3em}{0.8ex}\\
     \midrule
     Run malicious code (1)\cite{tseng2022dark} &\multirow{2}{*}{Cloud} &Remote code execution &Software &Cloud storage &\NumberCircle[red]{3} &\FilledCircle &\ProgressBar{0.87}{3em}{0.8ex}\\
     Unauthorized access (1)\cite{gulhane2019security} & &Spoofed identity & Social engineering &User account &\NumberCircle[red]{3} &\FilledCircle &\ProgressBar{0.87}{3em}{0.8ex}\\
    \bottomrule
  \end{tabular}
  \vspace{-0.1in}
\end{table*}

\begin{table*}[t]
  \centering
  \footnotesize
  \caption{An overview of XR Defense Approaches}
  \label{tab:defense-approaches}
  \begin{tabular}{@{} lll l cc l c @{}}
    \toprule
    \multirow{2}{*}{\textbf{Approach}}
      & \multirow{2}{*}{\textbf{Group}}
      & \multirow{2}{*}{\textbf{Mitigation}}
       & \multirow{2}{*}{\textbf{Defense Vector}}
      & \multicolumn{2}{c}{\textbf{Deployability}}
      & \multicolumn{2}{c}{\textbf{Robustness}}
      \\
    \cmidrule(lr){5-6} \cmidrule(lr){7-8}
      &  &  &  & \textbf{Trade-off} & \textbf{Maintenance} & \textbf{Efficacy}  &  \textbf{Stage}\\
      
    \midrule
    Keystroke inference \textcolor{green}{\cite{yang2024can}}
      & \multirow{8}{*}{Data Obfuscation} & Limit access to telemetry & Hand tracking API     
      & \EmptyCircle            & \NumberCircle[green]{1}        
      & \ProgressBar{0.53}{3em}{0.8ex}            & P\\
    
      Rate-limiting \textcolor{green}{\cite{slocum2023going}} &  &Reduced IMU sampling to 5Hz &Motion signals &\HalfFilledCircleLeft &\NumberCircle[green]{1} &\ProgressBar{0.4}{3em}{0.8ex} &P  \\
      Noise Addition \textcolor{green}{\cite{yang2024can}} &  &Adding zero-mean Gaussian noise &3D hand-tracking data &\EmptyCircle & \NumberCircle[green]{1} & \ProgressBar{0.46}{3em}{0.8ex} & P \\
     Randomized Keyboard\cite{wang2024gazeploit} & &Randomization of virtual keyboard &Feedback of keystrokes &\HalfFilledCircleLeft &\NumberCircle[yellow]{2} &\ProgressBar{0.6}{3em}{0.8ex} &P\\

     Avatars\cite{meng2024anonymizing} &   &Adding noise to gait data &Motion data &\EmptyCircle &\NumberCircle[green]{1} &\ProgressBar{0.7}{3em}{0.8ex} &P\\
     Privacy in motion\cite{sunprivacy} &  &Adding Laplace noise &Motion data stream &\HalfFilledCircleLeft &\NumberCircle[yellow]{2} &\ProgressBar{0.7}{3em}{0.8ex} &P\\
      Eye Tracking\cite{steil2019privacy} & &Using controlled noise &Sensor O/P layer &\HalfFilledCircleLeft &\NumberCircle[yellow]{2} &\ProgressBar{0.85}{3em}{0.8ex} &P\\
     Location Privacy\cite{shang2020arspy} &  &Reduce location precision & Network layer &\FilledCircle &\NumberCircle[yellow]{2} &\ProgressBar{0.85}{3em}{0.8ex} &P\\
\midrule
     
     Biometric Auth.\cite{wang2021nod,funk2019lookunlock} &\multirow{6}{*}{Authentication} &Head-neck movement modeling &IMU telemetry &\EmptyCircle &\NumberCircle[yellow]{2} &\ProgressBar{0.95}{3em}{0.8ex} &P\\
 Biometric anonymization\cite{mustafa2018unsure} & &Feature Suppression &Application Layer &\FilledCircle &\NumberCircle[yellow]{2} &\ProgressBar{0.7}{3em}{0.8ex} &P\\
Eye Tracking\cite{david2021privacy,bozkir2021differential} & &Gatekeeper API &Gaze data API  &\HalfFilledCircleLeft &\NumberCircle[yellow]{2} &\ProgressBar{0.75}{3em}{0.8ex} &P\\
     Shoulder Surfing\cite{duzgun2022shoulder} & &Graphical Password &Application layer &\HalfFilledCircleLeft &\NumberCircle[yellow]{2} &\ProgressBar{0.9}{3em}{0.8ex} &P\\
     EyeVEIL\cite{john2019eyeveil} & &Gaussian blur &Eye‐camera O/P &\HalfFilledCircleLeft &\NumberCircle[yellow]{2} &\ProgressBar{0.9}{3em}{0.8ex} &P\\
  Video Encryption\cite{hu2024exploring} &  &ROI video encryption & Video encoder  &\HalfFilledCircleLeft &\NumberCircle[yellow]{2} &\ProgressBar{0.8}{3em}{0.8ex} &P\\

  \midrule
      ShareAR\cite{ruth2019secure} &\multirow{7}{*}{Access Control} &Physical-world integration controls &App-level APIs &\EmptyCircle &\NumberCircle[yellow]{2} &\ProgressBar{0.85}{3em}{0.8ex} &P\\
    
     Privacy-Manager\cite{lehman2017privacymanager} &  &Context-aware policy &OS middleware &\HalfFilledCircleLeft &\NumberCircle[red]{3} &\ProgressBar{0.85}{3em}{0.8ex} &P\\
     
     Privacy Leakage\cite{wu2023privacy} & &Precision Reduction &Sensor API middleware &\EmptyCircle &\NumberCircle[yellow]{2} &\ProgressBar{0.5}{3em}{0.8ex} &P\\
    
     PrivXR\cite{warin2024privxr} & &Privacy panel &XR feature APIs &\HalfFilledCircleLeft &\NumberCircle[yellow]{2} &\ProgressBar{0.7}{3em}{0.8ex} &P\\
     
     Collaborative AR\cite{vallasciani2024handling} & &Salted-hash passwords &Application layer &\HalfFilledCircleLeft &\NumberCircle[yellow]{2} &\ProgressBar{0.9}{3em}{0.8ex} &P\\
     
     Cloth try-on\cite{sekhavat2016privacy} & &secure computation &Client-server data &\FilledCircle &\NumberCircle[yellow]{2} &\ProgressBar{0.76}{3em}{0.8ex} &P\\
     Knowledge\cite{mathis2020knowledge} & &PIN entry+biometric &Hand movement &\HalfFilledCircleLeft &\NumberCircle[yellow]{2} &\ProgressBar{0.98}{3em}{0.8ex} &P\\
    \bottomrule
  \end{tabular}
  \vspace{-0.1in}
\end{table*}
\vspace{-0.1in}
\section{XR-PRISM}\label{sec:xr-prism}
Our findings show that most XR side-channel and inference attacks require only minimal privileges (scores of 1–2) and modest expertise—few demand specialized hardware or deep reverse-engineering. For example, simply running a benign-looking XR app can grant access to rich motion or gaze streams, letting an attacker infer keystrokes, current application, or even bystander presence with off-the-shelf ML models\cite{gulhane2019security}. Healthcare XR deployments blend rich sensory inputs, real‐time rendering, haptic feedback, and sensitive biometric streams, creating intertwined security and privacy exposures. To quantify and prioritize these exposures, we extend a \textit{Multi‐Criteria Decision Analysis (MCDA)}\cite{linkov2011multi} based scoring framework to jointly assess both security and privacy risks. This approach consists of two main steps:\circledblack{1}
Scaling key parameters as per threat characteristics and \circledblack{2}Calculating the \textit{RiskScore} for taking immediate mitigation action. These key elements together form the \XRPRISM (XR-Privacy and Risk Impact Scoring Metric).
\begin{table}[ht]
  \centering
  \footnotesize
  \caption{Risk Assessment Model Parameters}
  \label{tab:risk-parameters}
  \begin{tabular}{@{}lclc@{}}
    \toprule
    \textbf{Risk Factor}  &\textbf{Symbol}    & \textbf{Description}   & \textbf{Weight($W$)} \\
    \midrule
    Threat Likelihood &($L$)   & Probability of attack        & 0.15 \\
    System Vulnerabilities &($V$) & Known XR platform flaws             & 0.15 \\
    Attack Surface &($A$)      & exposure of interfaces        & 0.10 \\
    Safety Impact &($I_s$)     & Patient-harm         & 0.30 \\
    Privacy Impact &($I_p$)    & Severity of user indentification           & 0.20 \\
    Control Effectiveness &($C$) & Strength of auth, encryption     & 0.10 \\
    \bottomrule
  \end{tabular}
\vspace{-0.2in}
\end{table}
\paragraph{\circled{1} \textbf{Key Risk Factors for Weighting}}
The proposed structure in \cite{ganin2020multicriteria} is intended to evaluate a cyber system’s risk using threats, vulnerabilities and consequences as the most significant criteria in order to choose the best remedial strategy. Expanding on their idea, the scoring mechanism developed
here scores each risk factor from 1 (low risk) to 10 (high
risk). We began by assigning weights to the six risk criterions, namely, \circledblack{1}Threat Likelihood, \circledblack{2}System Vulnerabilities, \circledblack{3}Attack Surface, \circledblack{4}Safety Impact, \circledblack{5} Privacy Impact and \circledblack{6} Control Effectiveness; shown in Table \ref{tab:risk-parameters}. The choice of weights reflects the specific regulatory and clinical sensitivities of XR healthcare systems. Following NIST SP 800-30 guidelines \cite{maclean2017nist,nist80030}for risk assessment and the approach in cyber risk modeling \cite{ganin2020multicriteria}, we prioritized patient safety ($W_{I_{s}}=0.30$) and privacy impact ($W_{I_{p}}=0.20$), since these dimensions are most closely tied to regulatory compliance and potential patient harm. Threat Likelihood ($W_L = 0.15$) and System Vulnerabilities ($W_V = 0.15$) were given medium weight, as they capture technical exploitability but do not always translate into direct clinical consequences \cite{spanakis2020cyber}. Attack Surface ($W_A = 0.10$) and Control Effectiveness ($W_C = 0.10$) were weighted lower, as these are primarily system-design factors that can vary widely across platforms \cite{pipyros2014cyber}. We treat these weights as initial heuristics; we plan a Delphi-style expert elicitation to empirically calibrate them. In practice, these weights would be obtained from XR security specialists using established procedures \cite{buede2024engineering} in an empirical implementation of this paradigm, depending on the attributes of the XR healthcare system. The weights add up to 1.00 to ensure probabilistic balanced scoring.

\paragraph{ \circled{2} \textbf{Formula and Scoring Interpretation}} Subsequent to defining and quantifying the parameters, the overall risk score is calculated using a weighted sum in Eq.\ref{eq:riskscore}. Control Effectiveness ($C$) is subtracted from 10 because stronger controls reduce risk.

We score six factors on a 1–10 scale and compute a single \emph{RiskScore} as a weighted sum:
\begin{equation}\label{eq:riskscore}
\mathit{RiskScore} = L\,W_{L} + V\,W_{V} + A\,W_{A} + I_{s}\,W_{I_{s}} + I_{p}\,W_{I_{p}} + (10 - C)\,W_{C}
\end{equation}
where:
\begin{itemize}
  \item $L$ (\emph{Threat Likelihood}): probability of an attack, informed by incident data and exploitability indices.
  \item $V$ (\emph{System Vulnerabilities}): count and severity of known flaws in firmware, runtime, and architecture.
  \item $A$ (\emph{Attack Surface}): number and exposure level of sensors, APIs, and network links
  \item $I_s$ (\emph{Safety Impact}): potential for patient harm (haptics, motion-sickness) 
  \item $I_p$ (\emph{Privacy Impact}): severity of PHI leakage or behavioral profiling 
  \item $C$ (\emph{Control Effectiveness}): strength of authentication, encryption, and session isolation (higher $C$ is more effective).
\end{itemize}
We choose weights to reflect the paramount importance of patient safety and data confidentiality:
\begin{equation}
\begin{aligned}
W_L &= 0.15,\quad W_V = 0.15,\quad W_A = 0.10,\\
W_{I_s} &= 0.30,\quad W_{I_p} = 0.20,\quad W_C = 0.10,
\end{aligned}
\end{equation}
with $\sum W=1.00$.
Using the formula, the risk score that is calculated is assigned Risk Levels from Low to Critical as per Table \ref{tab:risk-interpretation}. From the risk level, the required mitigation priority and appropriate action to be undertaken for the threat can be determined. An organization prioritizes outcomes and controls that can manage the risks with the most negative impacts and that are most cost-effective for their risk management results by using the
principles outlined in NIST SP 800-53: Security and Privacy Controls for Information Systems and Organizations \cite{force2013security}. Based on the principles outlined, our \XRPRISM can be utilized to interpret the required level of action for threat mitigation. \XRPRISM extends beyond CVSS and DREAD by explicitly folding in safety and privacy impacts—critical in healthcare XR—via two dedicated factors, \textit{Safety Impact} and \textit{Privacy Impact} each weighted heavily to reflect patient‐harm and PHI leakage concerns. Thereafter, we map the computed \textit{RiskScore} into actionable tiers:
\begin{table}[ht]
  \centering
  \footnotesize
  \caption{Risk Score Interpretation}
  \label{tab:risk-interpretation}
  \begin{tabular}{@{}ccc@{}}
    \toprule
    \textbf{RiskScore} & \textbf{Risk Level}  & \textbf{Mitigation Action}      \\
    \midrule
    $1.0$–$3.0$            & Low       & Monitor routinely;no immediate change  \\
    $3.1$–$6.0$            & Moderate  & Deploy preventive controls\\
    $6.1$–$8.0$            & High      & Immediate mitigation; elevate priority  \\
    $8.1$–$10.0$           & Critical  & Emergency response; consider system shutdown \\
    \bottomrule
  \end{tabular}
  \vspace{-0.1in}
\end{table}
\paragraph{Example.}
A tele‐therapy VR system suffers a motion‐replay attack that risks both user disorientation and PHI inference. Experts rate:
\begin{equation}\label{eq:values}
    L=6,\;V=5,\;A=7,\;I_s=8,\;I_p=9,\;C=4.
\end{equation}
\begin{equation}\label{eq:example}
  \mathit{RiskScore}
    =6\cdot0.15 + 5\cdot0.15 + 7\cdot0.10 
     + 8\cdot0.30 + 9\cdot0.20 + (10-4)\cdot0.10
    = 6.5
\end{equation}
placing it in the **High** tier.  We therefore recommend urgent deployment of signed telemetry, anomaly detection at the edge, and end‐to‐end encryption of all biometric streams.

By explicitly modeling both safety‐critical and privacy‐critical factors, \XRPRISM guides healthcare XR practitioners in data‐driven risk prioritization and balanced defense planning.
\begin{figure}[t]
  \centering
  \includegraphics[width=0.95\columnwidth]{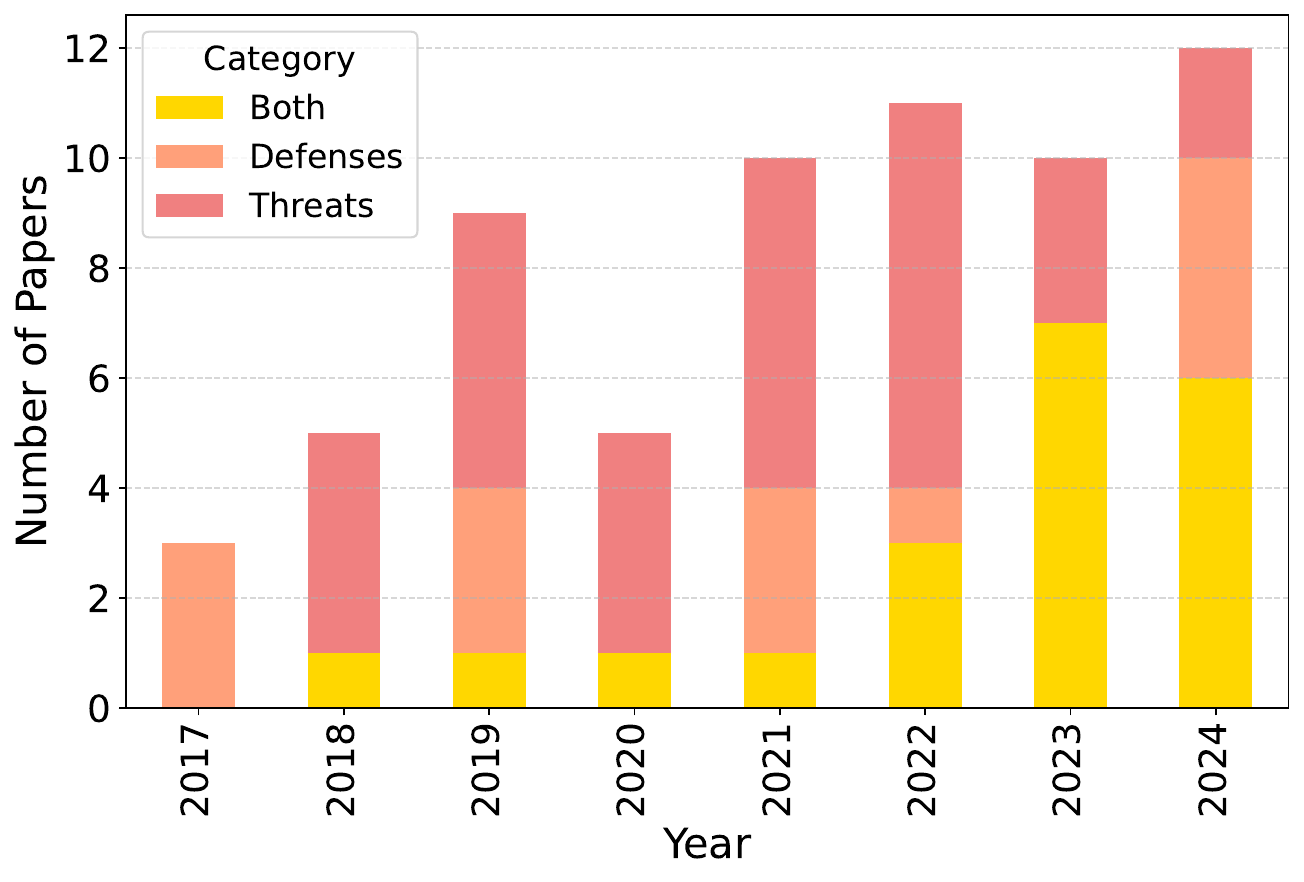}
  \caption{Publications by Year and Category }
  \label{fig:chart}
  \vspace{-0.2in}
\end{figure}
\section{Research Gaps}\label{sec:challenges}
By examining the information in Tables \ref{tab:attack-taxonomy} and \ref{tab:defense-approaches}, we now concentrate on determining research gaps and suggestions for further research.
\paragraph{\textbf{Low Prerequisites for Most Attacks}}
An analysis of Table \ref{tab:attack-taxonomy} illustrates that the majority of documented attack methods demand minimal prerequisites—most techniques score a 1 or 2 on the “Requisite” scale, indicating that an adversary needs little more than basic access (e.g., network proximity or unprotected APIs) to launch them. Only a handful \cite{slocum2023going,zhang2023s} reach a prerequisite level of 3. This skew toward low-barrier exploits suggests that XR systems are broadly exposed to attacks by relatively unsophisticated adversaries, underscoring the urgent need to harden default configurations and elevate baseline security measures.
\paragraph{\textbf{Lack of Shared Code and Reproducibility}}
Despite the variety of attack and defense strategies surveyed, there is scant evidence \cite{nair2023unique, ruth2019secure} of publicly released implementations. Very few papers accompany their contributions with open-source code or detailed experimental artifacts, making independent verification and comparative evaluation difficult. This gap not only hinders reproducibility—a cornerstone of scientific progress but also slows the community’s ability to build upon existing defenses. Establishing community repositories or requiring artifact disclosure (much as some security conferences like USENIX, ACM CCS, ACM WiSec now mandate) would greatly enhance transparency and accelerate innovation.
\paragraph{\textbf{Scarcity of Countermeasures}}
Although an increasing trend of S\&P publications can be seen over the years in Fig.\ref{fig:chart}, Table \ref{tab:defense-approaches} catalogs only about 20\% of total papers as countermeasures against the dozens of attacks in Table \ref{tab:attack-taxonomy}, revealing a pronounced mismatch between threat volume and mitigation effort evident. In particular, network and cloud layer defenses are underrepresented: whereas device or user layer controls are proposed, only a handful target higher-level threats such as session hijacking or remote code execution. This limited defensive breadth leaves many vectors, especially those affecting multi-user scenarios and backend services—effectively unprotected.
\paragraph{\textbf{Weaknesses in Existing Countermeasures}}
Even when countermeasures are proposed, their efficacy often remains low or modest. Most device-layer techniques (e.g., noise addition, rate limiting) trade off security gains against significant usability or performance costs, and their real-world robustness against adaptive adversaries is unclear. For example, randomized keyboards and Laplace-noise obfuscation can frustrate certain side-channel inferences but degrade user experience and may be bypassed by smarter attacks. Without rigorous, standardized evaluations of resilience under adversarial conditions, these defenses provide only a fragile first line of protection.
\paragraph{\textbf{Predominance of Preventative over Detective and Recovery Controls}}
Nearly all surveyed defenses are preventative—aimed at blocking attacks before they occur; yet there is a dearth of mechanisms for detection, forensics, or automated recovery post-compromise. The SoK itself notes that fewer than 30\% of proposals even report standardized risk or anomaly-detection metrics, and only 15\% include user-study–driven usability assessments. This lack of holistic, end-to-end security frameworks means that when prevention fails—as it inevitably will—systems lack the means to detect breaches, attribute them accurately, and restore safe operation.
\paragraph{\textbf{Underexplored Cloud-Layer Threats and Defenses}}
While cloud-based analytics and federated learning are increasingly central to XR healthcare, few studies examine attacks on these services (e.g., poisoning of federated models, inference from aggregated telemetry) or propose cloud-native defenses like homomorphic encryption or secure enclaves\cite{horbenko2025confidential}.
\paragraph{\textbf{Underexplored Healthcare Workflow}}
Despite a rapidly growing corpus of work on XR applications in healthcare—covering teletherapy, remote rehabilitation, surgical training, and beyond, very few of these studies evaluate or even acknowledge the security and privacy implications of their systems. Conversely, most XR security/privacy research is demonstrated in synthetic or generic testbeds, without any user studies or deployments in actual clinical workflows. This disconnect leaves us with unvalidated defenses whose real-world efficacy and usability in healthcare remain unknown.
\section{Related Work}\label{sec:related}
As far as we are aware, the study by Lake et al.\cite{lake2024cybersecurity} is the only scoping review that focuses on Extended Reality Healthcare applications attacks and defenses. But our work is distinct in three ways. Firstly, ours is a Systematization of Knowledge on in-depth security and privacy issues that exist in an XR threat model. Secondly, we formulate a more comprehensive approach to develop our attack and defense taxonomy. Lastly, we design a quantitative framework for risk scoring according to the needs of trending XR research.

\textit{Guzman et al.}\cite{de2019security} list and examine the various protection strategies that have been put out to guarantee the security and privacy of users and data in MR. They offer a data-driven classification of the different works into five main categories. \textit{Giaretta}\cite{giaretta2024security} examines the most recent developments in VR privacy and security, classifying possible problems and risks, and examines the origins and consequences of the threats that have been uncovered along with authentication in VR. The use of behavioral data as an identifying mechanism in immersive virtual reality applications raises privacy concerns, which are examined by \textit{Kumarapeli et al.}\cite{kumarapeli2024privacy}. However, these works partially focus on S\&P issues of VR/MR without considering the entire XR system pipeline for threat modeling and do not address the implications in terms of the healthcare industry.
\section{Conclusion}\label{sec:conclusion}
The capabilities of XR makes it an indispensable tool for safer, more effective, and more equitable healthcare delivery. The delivery of care and possibly the safety and well-being of people may be impacted if an attacker compromises a clinical VR device and tampers with the content and delivery of a clinical VR immersive session. In this paper, we provide the first comprehensive systematization of knowledge on XR privacy and security in healthcare, surveying 65 studies and organizing threats and defenses into a four-layer taxonomy (device, user, network and cloud) alongside a data-driven catalog of countermeasures; we introduce \XRPRISM, a weighted six-factor risk-scoring framework that unifies exploitability, safety and privacy impacts, and control effectiveness into a single actionable metric. We hope this data-driven study will help newcomers in this field as well as experts looking to broaden the scope of XR research.


\bibliographystyle{ACM-Reference-Format}
\bibliography{bibliography}

@String{Computing = "Computing" }

@String{Computer = "{IEEE} Computer" }

@String{Springer = "Springer-Verlag" }

@online{mitre_attack_ics,
  title   = {Matrix | MITRE ATT\&ACK\textregistered},
  author  = {{MITRE ATT\&CK}},
  year    = {2023},
  url     = {https://attack.mitre.org/matrices/ics/},
  urldate = {2025-04-10}
}

@techreport{FDA2023CybersecurityMedicalDevices,
  title        = {Cybersecurity in Medical Devices: Quality System Considerations and Content of Premarket Submissions},
  author       = {{U.S. Department of Health and Human Services, Food and Drug Administration}},
  institution  = {U.S. Food and Drug Administration},
  type         = {Guidance for Industry and Food and Drug Administration Staff},
  year         ={2023},
  url          = {https://www.fda.gov/media/119933/download},
}

@article{de2019security,
  title={Security and privacy approaches in mixed reality: A literature survey},
  author={De Guzman, Jaybie A and Thilakarathna, Kanchana and Seneviratne, Aruna},
  journal={ACM Computing Surveys (CSUR)},
  volume={52},
  number={6},
  pages={1--37},
  year={2019},
  publisher={ACM New York, NY, USA}
}

@article{giaretta2024security,
  title={Security and privacy in virtual reality: a literature survey},
  author={Giaretta, Alberto},
  journal={Virtual Reality},
  volume={29},
  number={1},
  pages={10},
  year={2024},
  publisher={Springer}
}

@inproceedings{xu2024review,
  title={A Review of Motion Data Privacy in Virtual Reality},
  author={Xu, Depeng and Wang, Weichao and Lu, Aidong},
  year={2024},
  organization={IEEE International Conference on Meta Computing}
}

@book{bell2022business,
  title={Business research methods},
  author={Bell, Emma and Harley, Bill and Bryman, Alan},
  year={2022},
  publisher={Oxford university press}
}

@book{creswell2016qualitative,
  title={Qualitative inquiry and research design: Choosing among five approaches},
  author={Creswell, John W and Poth, Cheryl N},
  year={2016},
  publisher={Sage publications}
}

@book{fink2019conducting,
  title={Conducting research literature reviews: From the internet to paper},
  author={Fink, Arlene},
  year={2019},
  publisher={Sage publications}
}

@article{okoli2015guide,
  title={A guide to conducting a systematic literature review of information systems research},
  author={Okoli, Chitu and Schabram, Kira},
  year={2015},
  publisher={Ssrn}
}

@article{ofte2023understanding,
  title={Understanding situation awareness in SOCs, a systematic literature review},
  author={Ofte, H{\aa}vard Jakobsen and Katsikas, Sokratis},
  journal={Computers \& Security},
  volume={126},
  pages={103069},
  year={2023},
  publisher={Elsevier}
}

@article{patel2022systematic,
  title={A systematic literature review on virtual reality and augmented reality in terms of privacy, authorization and data-leaks},
  author={Patel, Parth Dipakkumar and Trivedi, Prem},
  journal={arXiv preprint arXiv:2212.04621},
  year={2022}
}

@online{wikipedia_htcvive_2025,
  author   = {{Wikipedia contributors}},
  title    = {HTC Vive},
  year     = {2025},
  url      = {https://en.wikipedia.org/wiki/HTC_Vive},
  urldate  = {2025-02-19},
  
}

@inproceedings{yang2024can,
  title={Can virtual reality protect users from keystroke inference attacks?},
  author={Yang, Zhuolin and Sarwar, Zain and Hwang, Iris and Bhaskar, Ronik and Zhao, Ben Y and Zheng, Haitao},
  booktitle={33rd USENIX Security Symposium (USENIX Security 24)},
  pages={2725--2742},
  year={2024}
}

@inproceedings{zhang2023facereader,
  title={FaceReader: unobtrusively mining vital signs and vital sign embedded sensitive info via AR/VR motion sensors},
  author={Zhang, Tianfang and Ye, Zhengkun and Mahdad, Ahmed Tanvir and Akanda, Md Mojibur Rahman Redoy and Shi, Cong and Wang, Yan and Saxena, Nitesh and Chen, Yingying},
  booktitle={Proceedings of the 2023 ACM SIGSAC Conference on Computer and Communications Security},
  pages={446--459},
  year={2023}
}

@inproceedings{andrade2020discerning,
  title={Discerning user activity in extended reality through side-channel accelerometer observations},
  author={Andrade, Tiago Martins and Smith-Creasey, Max and Roscoe, Jonathan Francis},
  booktitle={2020 IEEE international conference on intelligence and security informatics (ISI)},
  pages={1--3},
  year={2020},
  organization={IEEE}
}

@inproceedings{slocum2023going,
  title={Going through the motions:$\{$AR/VR$\}$ keylogging from user head motions},
  author={Slocum, Carter and Zhang, Yicheng and Abu-Ghazaleh, Nael and Chen, Jiasi},
  booktitle={32nd USENIX Security Symposium (USENIX Security 23)},
  pages={159--174},
  year={2023}
}

@inproceedings{balzarotti2008clearshot,
  title={Clearshot: Eavesdropping on keyboard input from video},
  author={Balzarotti, Davide and Cova, Marco and Vigna, Giovanni},
  booktitle={2008 IEEE Symposium on Security and Privacy (sp 2008)},
  pages={170--183},
  year={2008},
  organization={IEEE}
}

@article{banerjee2012biometric,
  title={Biometric authentication and identification using keystroke dynamics: A survey},
  author={Banerjee, Salil P and Woodard, Damon L},
  journal={Journal of Pattern recognition research},
  volume={7},
  number={1},
  pages={116--139},
  year={2012},
  publisher={Citeseer}
}

@article{ferrer2021bias,
  title={Bias and discrimination in AI: a cross-disciplinary perspective},
  author={Ferrer, Xavier and Van Nuenen, Tom and Such, Jose M and Cot{\'e}, Mark and Criado, Natalia},
  journal={IEEE Technology and Society Magazine},
  volume={40},
  number={2},
  pages={72--80},
  year={2021},
  publisher={IEEE}
}

@article{zielinski2015exploring,
  title={Exploring the effects of image persistence in low frame rate virtual environments. In 2015 IEEE Virtual Reality (VR)},
  author={Zielinski, David J and Rao, Hrishikesh M and Sommer, Mark A and Kopper, Regis},
  journal={IEEE},
  volume={4},
  pages={19--26},
  year={2015}
}

@inproceedings{gopal2023hidden,
  title={Hidden reality: Caution, your hand gesture inputs in the immersive virtual world are visible to all!},
  author={Gopal, Sindhu Reddy Kalathur and Shukla, Diksha and Wheelock, James David and Saxena, Nitesh},
  booktitle={32nd USENIX security symposium (USENIX Security 23)},
  pages={859--876},
  year={2023}
}

@inproceedings{al2021vr,
  title={Vr-spy: A side-channel attack on virtual key-logging in vr headsets},
  author={Al Arafat, Abdullah and Guo, Zhishan and Awad, Amro},
  booktitle={2021 IEEE Virtual Reality and 3D User Interfaces (VR)},
  pages={564--572},
  year={2021},
  organization={IEEE}
}

@techreport{nist80030,
  author       = {{Joint Task Force Transformation Initiative}},
  title        = {Guide for Conducting Risk Assessments},
  institution  = {National Institute of Standards and Technology},
  number       = {SP 800-30 Rev.\ 1},
  year         = {2012},
  address      = {Gaithersburg, MD},
  url          = {https://doi.org/10.6028/NIST.SP.800-30r1}
}

@inproceedings{ling2019know,
  title={I know what you enter on gear vr},
  author={Ling, Zhen and Li, Zupei and Chen, Chen and Luo, Junzhou and Yu, Wei and Fu, Xinwen},
  booktitle={2019 IEEE Conference on Communications and Network Security (CNS)},
  pages={241--249},
  year={2019},
  organization={IEEE}
}

@inproceedings{shi2021face,
  title={Face-Mic: inferring live speech and speaker identity via subtle facial dynamics captured by AR/VR motion sensors},
  author={Shi, Cong and Xu, Xiangyu and Zhang, Tianfang and Walker, Payton and Wu, Yi and Liu, Jian and Saxena, Nitesh and Chen, Yingying and Yu, Jiadi},
  booktitle={Proceedings of the 27th Annual International Conference on Mobile Computing and Networking},
  pages={478--490},
  year={2021}
}

@inproceedings{wang2024gazeploit,
  title={GAZEploit: Remote Keystroke Inference Attack by Gaze Estimation from Avatar Views in VR/MR Devices},
  author={Wang, Hanqiu and Zhan, Zihao and Shan, Haoqi and Dai, Siqi and Panoff, Maximilian and Wang, Shuo},
  booktitle={Proceedings of the 2024 on ACM SIGSAC Conference on Computer and Communications Security},
  pages={1731--1745},
  year={2024}
}

@inproceedings{nair2023unique,
  title={Unique identification of 50,000+ virtual reality users from head \& hand motion data},
  author={Nair, Vivek and Guo, Wenbo and Mattern, Justus and Wang, Rui and O'Brien, James F and Rosenberg, Louis and Song, Dawn},
  booktitle={32nd USENIX Security Symposium (USENIX Security 23)},
  pages={895--910},
  year={2023}
}

@online{apple_vision_pro,
  author   = {{Apple Inc.}},
  title    = {Apple Vision Pro},
  year     = {2025},
  url      = {https://www.apple.com/apple-vision-pro/},
  urldate  = {2025-07-05},
  note     = {Accessed: 2025-07-03}
}

@article{jiang2022learning,
  title={Learning to type with mobile keyboards: Findings with a randomized keyboard},
  author={Jiang, Xinhui and Jokinen, Jussi PP and Oulasvirta, Antti and Ren, Xiangshi},
  journal={Computers in Human Behavior},
  volume={126},
  pages={106992},
  year={2022},
  publisher={Elsevier}
}

@inproceedings{khalili2024virtual,
  title={Virtual Keymysteries Unveiled: Detecting Keystrokes in VR with External Side-Channels},
  author={Khalili, Hossein and Chen, Alexander and Papaiakovou, Theodoros and Jacques, Timothy and Chien, Hao-Jen and Liu, Changwei and Ding, Aolin and Hass, Amin and Zonouz, Saman and Sehatbakhsh, Nader},
  booktitle={2024 IEEE Security and Privacy Workshops (SPW)},
  pages={260--266},
  year={2024},
  organization={IEEE}
}

@inproceedings{ruth2019secure,
  title={Secure $\{$Multi-User$\}$ content sharing for augmented reality applications},
  author={Ruth, Kimberly and Kohno, Tadayoshi and Roesner, Franziska},
  booktitle={28th USENIX Security Symposium (USENIX Security 19)},
  pages={141--158},
  year={2019}
}

@inproceedings{luo2022holologger,
  title={Holologger: Keystroke inference on mixed reality head mounted displays},
  author={Luo, Shiqing and Hu, Xinyu and Yan, Zhisheng},
  booktitle={2022 IEEE Conference on Virtual Reality and 3D User Interfaces (VR)},
  pages={445--454},
  year={2022},
  organization={IEEE}
}

@article{tricomi2023you,
  title={You can’t hide behind your headset: User profiling in augmented and virtual reality},
  author={Tricomi, Pier Paolo and Nenna, Federica and Pajola, Luca and Conti, Mauro and Gamberini, Luciano},
  journal={IEEE Access},
  volume={11},
  pages={9859--9875},
  year={2023},
  publisher={IEEE}
}

@article{meng2024anonymizing,
  title={De-Anonymizing Avatars in Virtual Reality: Attacks and Countermeasures},
  author={Meng, Yan and Zhan, Yuxia and Li, Jiachun and Du, Suguo and Zhu, Haojin and Shen, Xuemin},
  journal={IEEE Transactions on Mobile Computing},
  year={2024},
  publisher={IEEE}
}

@article{sunprivacy,
  title={Privacy in Motion: Implementing Differential Privacy for User Motion in VR},
  author={SUN, RUOXI and WANG, HANWEN and XUE, MINHUI and CHEN, HSIANG-TING}
}

@inproceedings{wang2021nod,
  title={Nod to auth: Fluent ar/vr authentication with user head-neck modeling},
  author={Wang, Xue and Zhang, Yang},
  booktitle={Extended Abstracts of the 2021 CHI Conference on Human Factors in Computing Systems},
  pages={1--7},
  year={2021}
}

@inproceedings{steil2019privacy,
  title={Privacy-aware eye tracking using differential privacy},
  author={Steil, Julian and Hagestedt, Inken and Huang, Michael Xuelin and Bulling, Andreas},
  booktitle={Proceedings of the 11th ACM Symposium on Eye Tracking Research \& Applications},
  pages={1--9},
  year={2019}
}

@inproceedings{valluripally2020attack,
  title={Attack trees for security and privacy in social virtual reality learning environments},
  author={Valluripally, Samaikya and Gulhane, Aniket and Mitra, Reshmi and Hoque, Khaza Anuarul and Calyam, Prasad},
  booktitle={2020 IEEE 17th Annual Consumer Communications \& Networking Conference (CCNC)},
  pages={1--9},
  year={2020},
  organization={IEEE}
}

@article{casey2019immersive,
  title={Immersive virtual reality attacks and the human joystick},
  author={Casey, Peter and Baggili, Ibrahim and Yarramreddy, Ananya},
  journal={IEEE Transactions on Dependable and Secure Computing},
  volume={18},
  number={2},
  pages={550--562},
  year={2019},
  publisher={IEEE}
}

@article{valluripally2021modeling,
  title={Modeling and defense of social virtual reality attacks inducing cybersickness},
  author={Valluripally, Samaikya and Gulhane, Aniket and Hoque, Khaza Anuarul and Calyam, Prasad},
  journal={IEEE Transactions on Dependable and Secure Computing},
  volume={19},
  number={6},
  pages={4127--4144},
  year={2021},
  publisher={IEEE}
}

@inproceedings{adams2018ethics,
  title={Ethics emerging: the story of privacy and security perceptions in virtual reality},
  author={Adams, Devon and Bah, Alseny and Barwulor, Catherine and Musaby, Nureli and Pitkin, Kadeem and Redmiles, Elissa M},
  booktitle={Fourteenth symposium on usable privacy and security (SOUPS 2018)},
  pages={427--442},
  year={2018}
}

@inproceedings{tseng2022dark,
  title={The dark side of perceptual manipulations in virtual reality},
  author={Tseng, Wen-Jie and Bonnail, Elise and McGill, Mark and Khamis, Mohamed and Lecolinet, Eric and Huron, Samuel and Gugenheimer, Jan},
  booktitle={Proceedings of the 2022 CHI Conference on Human Factors in Computing Systems},
  pages={1--15},
  year={2022}
}

@inproceedings{yarramreddy2018forensic,
  title={Forensic analysis of immersive virtual reality social applications: a primary account},
  author={Yarramreddy, Ananya and Gromkowski, Peter and Baggili, Ibrahim},
  booktitle={2018 IEEE Security and Privacy Workshops (SPW)},
  pages={186--196},
  year={2018},
  organization={IEEE}
}

@article{rafique2020tracking,
  title={Tracking attacks on virtual reality systems},
  author={Rafique, Muhammad Usman and Sen-ching, S Cheung},
  journal={IEEE Consumer Electronics Magazine},
  volume={9},
  number={2},
  pages={41--46},
  year={2020},
  publisher={IEEE}
}

@inproceedings{gulhane2019security,
  title={Security, privacy and safety risk assessment for virtual reality learning environment applications},
  author={Gulhane, Aniket and Vyas, Akhil and Mitra, Reshmi and Oruche, Roland and Hoefer, Gabriela and Valluripally, Samaikya and Calyam, Prasad and Hoque, Khaza Anuarul},
  booktitle={2019 16th IEEE annual consumer communications \& networking conference (CCNC)},
  pages={1--9},
  year={2019},
  organization={IEEE}
}

@inproceedings{lehman2017privacymanager,
  title={PrivacyManager: An access control framework for mobile augmented reality applications},
  author={Lehman, Sarah M and Tan, Chiu C},
  booktitle={2017 IEEE Conference on Communications and Network Security (CNS)},
  pages={1--9},
  year={2017},
  organization={IEEE}
}

@article{shang2020arspy,
  title={ARSpy: Breaking location-based multi-player augmented reality application for user location tracking},
  author={Shang, Jiacheng and Chen, Si and Wu, Jie and Yin, Shu},
  journal={IEEE Transactions on Mobile Computing},
  volume={21},
  number={2},
  pages={433--447},
  year={2020},
  publisher={IEEE}
}

@inproceedings{wu2023privacy,
  title={Privacy leakage via unrestricted motion-position sensors in the age of virtual reality: A study of snooping typed input on virtual keyboards},
  author={Wu, Yi and Shi, Cong and Zhang, Tianfang and Walker, Payton and Liu, Jian and Saxena, Nitesh and Chen, Yingying},
  booktitle={2023 IEEE Symposium on Security and Privacy (SP)},
  pages={3382--3398},
  year={2023},
  organization={IEEE}
}

@inproceedings{moore2021personal,
  title={Personal identifiability of user tracking data during VR training},
  author={Moore, Alec G and McMahan, Ryan P and Dong, Hailiang and Ruozzi, Nicholas},
  booktitle={2021 IEEE Conference on Virtual Reality and 3D User Interfaces Abstracts and Workshops (VRW)},
  pages={556--557},
  year={2021},
  organization={IEEE}
}

@article{nair2023truth,
  title={Truth in motion: The unprecedented risks and opportunities of extended reality motion data},
  author={Nair, Vivek and Rosenberg, Louis and O’Brien, James F and Song, Dawn},
  journal={IEEE Security \& Privacy},
  volume={22},
  number={1},
  pages={24--32},
  year={2023},
  publisher={IEEE}
}

@inproceedings{mustafa2018unsure,
  title={Unsure how to authenticate on your vr headset? come on, use your head!},
  author={Mustafa, Tahrima and Matovu, Richard and Serwadda, Abdul and Muirhead, Nicholas},
  booktitle={Proceedings of the Fourth ACM International Workshop on Security and Privacy Analytics},
  pages={23--30},
  year={2018}
}

@article{nair2023inferring,
  title={Inferring private personal attributes of virtual reality users from head and hand motion data},
  author={Nair, Vivek and Rack, Christian and Guo, Wenbo and Wang, Rui and Li, Shuixian and Huang, Brandon and Cull, Atticus and O'Brien, James F and Latoschik, Marc and Rosenberg, Louis and others},
  journal={arXiv preprint arXiv:2305.19198},
  year={2023}
}

@inproceedings{slocum2024doesn,
  title={That Doesn't Go There: Attacks on Shared State in $\{$Multi-User$\}$ Augmented Reality Applications},
  author={Slocum, Carter and Zhang, Yicheng and Shayegani, Erfan and Zaree, Pedram and Abu-Ghazaleh, Nael and Chen, Jiasi},
  booktitle={33rd USENIX Security Symposium (USENIX Security 24)},
  pages={2761--2778},
  year={2024}
}

@inproceedings{pfeuffer2019behavioural,
  title={Behavioural biometrics in vr: Identifying people from body motion and relations in virtual reality},
  author={Pfeuffer, Ken and Geiger, Matthias J and Prange, Sarah and Mecke, Lukas and Buschek, Daniel and Alt, Florian},
  booktitle={Proceedings of the 2019 CHI Conference on Human Factors in Computing Systems},
  pages={1--12},
  year={2019}
}

@inproceedings{zhang2023s,
  title={It's all in your head (set): Side-channel attacks on $\{$AR/VR$\}$ systems},
  author={Zhang, Yicheng and Slocum, Carter and Chen, Jiasi and Abu-Ghazaleh, Nael},
  booktitle={32nd USENIX Security Symposium (USENIX Security 23)},
  pages={3979--3996},
  year={2023}
}

@inproceedings{grichi2024biosensor,
  title={Biosensor-Instrumented xR Headsets: A Double-Edged Sword for User Identity and Privacy Management in the Metaverse},
  author={Grichi, Ihs{\^a}n and Jaberi, Mina and Falk, Tiago H},
  booktitle={2024 IEEE International Symposium on Mixed and Augmented Reality Adjunct (ISMAR-Adjunct)},
  pages={17--19},
  year={2024},
  organization={IEEE}
}

@inproceedings{warin2024privxr,
  title={PrivXR: A cross-platform privacy-preserving API and privacy panel for extended reality},
  author={Warin, Chris and Seeger, Dominik and Shams, Shirin and Reinhardt, Delphine},
  booktitle={2024 IEEE International Conference on Pervasive Computing and Communications Workshops and other Affiliated Events (PerCom Workshops)},
  pages={417--420},
  year={2024},
  organization={IEEE}
}

@article{david2021privacy,
  title={A privacy-preserving approach to streaming eye-tracking data},
  author={David-John, Brendan and Hosfelt, Diane and Butler, Kevin and Jain, Eakta},
  journal={IEEE Transactions on Visualization and Computer Graphics},
  volume={27},
  number={5},
  pages={2555--2565},
  year={2021},
  publisher={IEEE}
}

@inproceedings{liebers2021understanding,
  title={Understanding user identification in virtual reality through behavioral biometrics and the effect of body normalization},
  author={Liebers, Jonathan and Abdelaziz, Mark and Mecke, Lukas and Saad, Alia and Auda, Jonas and Gruenefeld, Uwe and Alt, Florian and Schneegass, Stefan},
  booktitle={Proceedings of the 2021 CHI Conference on Human Factors in Computing Systems},
  pages={1--11},
  year={2021}
}

@inproceedings{lehman2020stealthy,
  title={Stealthy privacy attacks against mobile ar apps},
  author={Lehman, Sarah M and Alrumayh, Abrar S and Ling, Haibin and Tan, Chiu C},
  booktitle={2020 IEEE Conference on Communications and Network Security (CNS)},
  pages={1--5},
  year={2020},
  organization={IEEE}
}

@inproceedings{duzgun2022shoulder,
  title={Shoulder-surfing resistant authentication for augmented reality},
  author={D{\"u}zg{\"u}n, Reyhan and Mayer, Peter and Volkamer, Melanie},
  booktitle={Nordic Human-Computer Interaction Conference},
  pages={1--13},
  year={2022}
}

@inproceedings{abdrabou2022understanding,
  title={Understanding shoulder surfer behavior and attack patterns using virtual reality},
  author={Abdrabou, Yasmeen and Rivu, Sheikh Radiah and Ammar, Tarek and Liebers, Jonathan and Saad, Alia and Liebers, Carina and Gruenefeld, Uwe and Knierim, Pascal and Khamis, Mohamed and Makela, Ville and others},
  booktitle={Proceedings of the 2022 International Conference on Advanced Visual Interfaces},
  pages={1--9},
  year={2022}
}

@inproceedings{john2019eyeveil,
  title={EyeVEIL: degrading iris authentication in eye tracking headsets},
  author={John, Brendan and Koppal, Sanjeev and Jain, Eakta},
  booktitle={Proceedings of the 11th ACM Symposium on Eye Tracking Research \& Applications},
  pages={1--5},
  year={2019}
}

@inproceedings{nguyen2024penetration,
  title={Penetration vision through virtual reality headsets: identifying 360-degree videos from head movements},
  author={Nguyen, Anh and Zhang, Xiaokuan and Yan, Zhisheng},
  booktitle={33rd USENIX Security Symposium (USENIX Security 24)},
  pages={2779--2796},
  year={2024}
}

@inproceedings{liebers2021using,
  title={Using gaze behavior and head orientation for implicit identification in virtual reality},
  author={Liebers, Jonathan and Horn, Patrick and Burschik, Christian and Gruenefeld, Uwe and Schneegass, Stefan},
  booktitle={Proceedings of the 27th ACM Symposium on Virtual Reality Software and Technology},
  pages={1--9},
  year={2021}
}

@inproceedings{vallasciani2024handling,
  title={Handling privacy and security aspects in a collaborative AR session},
  author={Vallasciani, Giacomo and Schinoppi, Andrea and Cascarano, Pasquale and Marfia, Gustavo and Donatiello, Lorenzo},
  booktitle={2024 IEEE International Symposium on Mixed and Augmented Reality Adjunct (ISMAR-Adjunct)},
  pages={20--22},
  year={2024},
  organization={IEEE}
}

@inproceedings{hu2024exploring,
  title={Exploring Device-Oriented Video Encryption for Hierarchical Privacy Protection in AR Content Sharing},
  author={Hu, Yongquan and Zheng, Dongsheng and Nie, Kexin and Zhang, Junyan and Hu, Wen and Quigley, Aaron},
  booktitle={2024 IEEE International Symposium on Mixed and Augmented Reality Adjunct (ISMAR-Adjunct)},
  pages={427--428},
  year={2024},
  organization={IEEE}
}

@article{andrews2019extended,
  title={Extended reality in medical practice},
  author={Andrews, Christopher and Southworth, Michael K and Silva, Jennifer NA and Silva, Jonathan R},
  journal={Current treatment options in cardiovascular medicine},
  volume={21},
  pages={1--12},
  year={2019},
  publisher={Springer}
}

@inproceedings{abraham2022implications,
  title={Implications of xr on privacy, security and behaviour: Insights from experts},
  author={Abraham, Melvin and Saeghe, Pejman and Mcgill, Mark and Khamis, Mohamed},
  booktitle={Nordic Human-Computer Interaction Conference},
  pages={1--12},
  year={2022}
}

@article{alhakamy2024extended,
  title={Extended reality (XR) toward building immersive solutions: the key to unlocking industry 4.0},
  author={Alhakamy, A’aeshah},
  journal={ACM Computing Surveys},
  volume={56},
  number={9},
  pages={1--38},
  year={2024},
  publisher={ACM New York, NY}
}

@inproceedings{sivelle2024extended,
  title={Extended Reality in critical sectors: Exploring the use cases and challenges},
  author={Sivelle, Camille and Palma, David and De Moor, Katrien},
  booktitle={Norsk IKT-konferanse for forskning og utdanning},
  number={2},
  year={2024}
}

@article{han2022comic,
  title={CoMIC: A collaborative mobile immersive computing infrastructure for conducting multi-user XR research},
  author={Han, Bo and Pathak, Parth and Chen, Songqing and Yu, Lap-Fai},
  journal={IEEE Network},
  volume={37},
  number={6},
  pages={124--131},
  year={2022},
  publisher={IEEE}
}

@inproceedings{rudzki2022xr,
  title={XR-based HRTF measurements},
  author={Rudzki, Tomasz and Murphy, Damian and Kearney, Gavin},
  booktitle={Audio Engineering Society Conference: 2022 AES International Conference on Audio for Virtual and Augmented Reality},
  year={2022},
  organization={Audio Engineering Society}
}

@article{lim2024impact,
  title={Impact of Mixed Reality-Based Rehabilitation on Muscle Activity in Lower-Limb Amputees: An EMG Analysis},
  author={Lim, Gyubeom and Youn, Heejun and Kim, Hyojin and Jeong, Hyunghwa and Cho, Jeongmok and Lee, Seunghyun and Pak, Changsik and Kwon, Soonchul},
  journal={IEEE Access},
  year={2024},
  publisher={IEEE}
}

@article{sheng2024review,
  title={Review on SLAM algorithms for Augmented Reality},
  author={Sheng, Xingdong and Mao, Shijie and Yan, Yichao and Yang, Xiaokang},
  journal={Displays},
  pages={102806},
  year={2024},
  publisher={Elsevier}
}

@article{shi2024impact,
  title={Impact of Extended Reality Applications on Communication Protocol Design},
  author={Shi, Yuying},
  year={2024}
}

@inproceedings{park2022instantxr,
  title={InstantXR: Instant XR environment on the web using hybrid rendering of cloud-based NeRF with 3d assets},
  author={Park, Moonsik and Yoo, Byounghyun and Moon, Jee Young and Seo, Ji Hyun},
  booktitle={Proceedings of the 27th International Conference on 3D Web Technology},
  pages={1--9},
  year={2022}
}

@article{huang2025model,
  title={Model between Network Quality and XR User’s Experience in Terms of Video Stalling for Cloud Rendering XR},
  author={Huang, Yuhong and Cao, Lei and Zhang, Hongbiao and Kong, Luting and Xu, Jingwen and Jin, Chenguang and Wang, Xintai},
  journal={IEEE Access},
  year={2025},
  publisher={IEEE}
}

@online{hhs_hipaa_1996,
  author   = {{U.S. Department of Health and Human Services}},
  title    = {Health Insurance Portability and Accountability Act of 1996 (HIPAA)},
  year     = {1996},
  url      = {https://www.hhs.gov/hipaa/index.html},
  urldate  = {2025-03-10},
  note     = {Public Law 104–191}
}

@misc{eu_gdpr,
  title        = {Regulation (EU) 2016/679 of the European Parliament and of the Council},
  howpublished = {Official Journal of the European Union, L\,119, 1--88},
  date         = {2016-05-04},
  url          = {https://eur-lex.europa.eu/eli/reg/2016/679/oj},
  urldate      = {2025-03-10}
}

@inproceedings{sofiadin2024threats,
  title={Threats of Extended Reality (XR) Applications to Teaching and Learning: Instructors’ perspectives.},
  author={Sofiadin, Aidrina},
  booktitle={2024 IEEE 25th International Symposium on a World of Wireless, Mobile and Multimedia Networks (WoWMoM)},
  pages={76--78},
  year={2024},
  organization={IEEE}
}

@book{pappas2021extended,
  title={Extended reality (xr) \& gamification in the context of the internet of things (iot) and artificial intelligence (ai)},
  author={Pappas, Georgios},
  year={2021},
  publisher={Michigan State University}
}

@techreport{first2019cvss,
  author       = {{Forum of Incident Response and Security Teams (FIRST)}},
  title        = {Common Vulnerability Scoring System Version 3.1: Specification Document},
  institution  = {FIRST},
  year         = {2019},
  url          = {https://www.first.org/cvss/v3-1/specification-document},
}

@misc{wikidread,
  author       = {{Wikipedia contributors}},
  title        = {DREAD (risk assessment model)},
 url = {https://en.wikipedia.org/wiki/DREAD_(risk_assessment_model)},
  year         = {2025},
  note         = {Accessed: 2025-03-10},
}

@inproceedings{koch2023ok,
  title={The $\{$OK$\}$ is not enough: A large scale study of consent dialogs in smartphone applications},
  author={Koch, Simon and Altpeter, Benjamin and Johns, Martin},
  booktitle={32nd USENIX Security Symposium (USENIX Security 23)},
  pages={5467--5484},
  year={2023}
}

@book{linkov2011multi,
  title={Multi-criteria decision analysis: environmental applications and case studies},
  author={Linkov, Igor and Moberg, Emily},
  year={2011},
  publisher={CRC Press}
}

@article{ganin2020multicriteria,
  title={Multicriteria decision framework for cybersecurity risk assessment and management},
  author={Ganin, Alexander A and Quach, Phuoc and Panwar, Mahesh and Collier, Zachary A and Keisler, Jeffrey M and Marchese, Dayton and Linkov, Igor},
  journal={Risk Analysis},
  volume={40},
  number={1},
  pages={183--199},
  year={2020},
  publisher={Wiley Online Library}
}

@book{buede2024engineering,
  title={The engineering design of systems: models and methods},
  author={Buede, Dennis M and Miller, William D},
  year={2024},
  publisher={John Wiley \& Sons}
}

@article{maclean2017nist,
  title={The NIST risk management framework: Problems and recommendations},
  author={Maclean, Don},
  journal={Cyber Security: A Peer-Reviewed Journal},
  volume={1},
  number={3},
  pages={207--217},
  year={2017},
  publisher={Henry Stewart Publications}
}

@article{force2013security,
  title={Security and privacy controls for federal information systems and organizations},
  author={Force, Joint Task and Initiative, Transformation},
  journal={NIST Special Publication},
  volume={800},
  number={53},
  pages={8--13},
  year={2013}
}

@article{lake2024cybersecurity,
  title={Cybersecurity and Privacy Issues in Extended Reality Health Care Applications: Scoping Review},
  author={Lake, Kaitlyn and Mc Kittrick, Andrea and Desselle, Mathilde and Bo, Antonio Padilha Lanari and Abayasiri, R Achintha M and Fleming, Jennifer and Baghaei, Nilufar and Kim, Dan Dongseong and others},
  journal={JMIR XR and Spatial Computing (JMXR)},
  volume={1},
  number={1},
  pages={e59409},
  year={2024},
  publisher={JMIR Publications Inc., Toronto, Canada}
}

@article{kumarapeli2024privacy,
  title={Privacy threats of behaviour identity detection in VR},
  author={Kumarapeli, Dilshani and Jung, Sungchul and Lindeman, Robert W},
  journal={Frontiers in Virtual Reality},
  volume={5},
  pages={1197547},
  year={2024},
  publisher={Frontiers Media SA}
}

@inproceedings{meteriz2022keylogging,
  title={A keylogging inference attack on air-tapping keyboards in virtual environments},
  author={Meteriz-Y{\i}ld{\i}ran, {\"U}lk{\"u} and Y{\i}ld{\i}ran, Necip Faz{\i}l and Awad, Amro and Mohaisen, David},
  booktitle={2022 IEEE Conference on Virtual Reality and 3D User Interfaces (VR)},
  pages={765--774},
  year={2022},
  organization={IEEE}
}

@article{shukla2019stealing,
  title={Stealing passwords by observing hands movement},
  author={Shukla, Diksha and Phoha, Vir V},
  journal={IEEE Transactions on Information Forensics and Security},
  volume={14},
  number={12},
  pages={3086--3101},
  year={2019},
  publisher={IEEE}
}

@article{bozkir2021differential,
  title={Differential privacy for eye tracking with temporal correlations},
  author={Bozkir, Efe and G{\"u}nl{\"u}, Onur and Fuhl, Wolfgang and Schaefer, Rafael F and Kasneci, Enkelejda},
  journal={Plos one},
  volume={16},
  number={8},
  pages={e0255979},
  year={2021},
  publisher={Public Library of Science San Francisco, CA USA}
}

@article{olade2020biomove,
  title={Biomove: Biometric user identification from human kinesiological movements for virtual reality systems},
  author={Olade, Ilesanmi and Fleming, Charles and Liang, Hai-Ning},
  journal={Sensors},
  volume={20},
  number={10},
  pages={2944},
  year={2020},
  publisher={MDPI}
}

@inproceedings{trimananda2022ovrseen,
  title={$\{$OVRseen$\}$: Auditing network traffic and privacy policies in oculus $\{$VR$\}$},
  author={Trimananda, Rahmadi and Le, Hieu and Cui, Hao and Ho, Janice Tran and Shuba, Anastasia and Markopoulou, Athina},
  booktitle={31st USENIX security symposium (USENIX security 22)},
  pages={3789--3806},
  year={2022}
}

@inproceedings{chen2018eyetell,
  title={Eyetell: Video-assisted touchscreen keystroke inference from eye movements},
  author={Chen, Yimin and Li, Tao and Zhang, Rui and Zhang, Yanchao and Hedgpeth, Terri},
  booktitle={2018 IEEE Symposium on Security and Privacy (SP)},
  pages={144--160},
  year={2018},
  organization={IEEE}
}

@inproceedings{funk2019lookunlock,
  title={Lookunlock: Using spatial-targets for user-authentication on hmds},
  author={Funk, Markus and Marky, Karola and Mizutani, Iori and Kritzler, Mareike and Mayer, Simon and Michahelles, Florian},
  booktitle={Extended Abstracts of the 2019 CHI Conference on Human Factors in Computing Systems},
  pages={1--6},
  year={2019}
}

@inproceedings{mathis2020knowledge,
  title={Knowledge-driven biometric authentication in virtual reality},
  author={Mathis, Florian and Fawaz, Hassan Ismail and Khamis, Mohamed},
  booktitle={Extended abstracts of the 2020 CHI conference on human factors in computing systems},
  pages={1--10},
  year={2020}
}

@article{fu2018writing,
  title={Writing in the air with WiFi signals for virtual reality devices},
  author={Fu, Zhangjie and Xu, Jiashuang and Zhu, Zhuangdi and Liu, Alex X and Sun, Xingming},
  journal={IEEE Transactions on Mobile Computing},
  volume={18},
  number={2},
  pages={473--484},
  year={2018},
  publisher={IEEE}
}

@inproceedings{miller2022combining,
  title={Combining real-world constraints on user behavior with deep neural networks for virtual reality (vr) biometrics},
  author={Miller, Robert and Banerjee, Natasha Kholgade and Banerjee, Sean},
  booktitle={2022 IEEE conference on virtual reality and 3D user interfaces (VR)},
  pages={409--418},
  year={2022},
  organization={IEEE}
}

@article{chen2022extended,
  title={Extended Reality (XR) and telehealth interventions for children or adolescents with autism spectrum disorder: Systematic review of qualitative and quantitative studies},
  author={Chen, Yuhan and Zhou, Zhuoren and Cao, Min and Liu, Min and Lin, Zhihao and Yang, Weixin and Yang, Xiao and Dhaidhai, Denzel and Xiong, Peng},
  journal={Neuroscience \& Biobehavioral Reviews},
  volume={138},
  pages={104683},
  year={2022},
  publisher={Elsevier}
}

@article{yang2021utilization,
  title={Utilization exercise rehabilitation using metaverse (VR{\textperiodcentered} AR{\textperiodcentered} MR{\textperiodcentered} XR)},
  author={Yang, Jeong Ok and Lee, Jook Sook},
  journal={Korean Journal of Applied Biomechanics},
  volume={31},
  number={4},
  pages={249--258},
  year={2021},
  publisher={Korean Society of Applied Biomechanics}
}

@article{worlikar2023mixed,
  title={Mixed reality platforms in telehealth delivery: scoping review},
  author={Worlikar, Hemendra and Coleman, Sean and Kelly, Jack and O’Connor, Sadhbh and Murray, Aoife and McVeigh, Terri and Doran, Jennifer and McCabe, Ian and O'Keeffe, Derek},
  journal={JMIR biomedical engineering},
  volume={8},
  pages={e42709},
  year={2023},
  publisher={JMIR Publications Toronto, Canada}
}

@article{lamb2025effectiveness,
  title={Effectiveness of XR-Based Exposure Therapy for Phobic Disorders},
  author={Lamb, Richard and Perry, Jason and Sutherland, Emily C and Hoston Jr, Douglas and Garris, Alex and DeRiggs, Aniya M},
  journal={Journal of Counseling \& Development},
  volume={103},
  number={3},
  pages={316--333},
  year={2025},
  publisher={Wiley Online Library}
}

@article{sekhavat2016privacy,
  title={Privacy preserving cloth try-on using mobile augmented reality},
  author={Sekhavat, Yoones A},
  journal={IEEE Transactions on Multimedia},
  volume={19},
  number={5},
  pages={1041--1049},
  year={2016},
  publisher={IEEE}
}

@article{horbenko2025confidential,
  title={Confidential Computing in Front-End: Enhancing Data Security with Secure Enclaves and Homomorphic Encryption},
  author={Horbenko, Yuliia},
  journal={International Journal of Advanced Multidisciplinary Research and Studies},
  volume={5},
  number={3},
  pages={308--321},
  year={2025}
}

@article{mattei2017privacy,
  title={Privacy, confidentiality, and security of health care information: Lessons from the recent Wannacry cyberattack},
  author={Mattei, Tobias A},
  journal={World neurosurgery},
  volume={104},
  pages={972--974},
  year={2017},
  publisher={Elsevier}
}

@inproceedings{spanakis2020cyber,
  title={Cyber-attacks and threats for healthcare--a multi-layer thread analysis},
  author={Spanakis, Emmanouil G and Bonomi, Silvia and Sfakianakis, Stelios and Santucci, Giuseppe and Lenti, Simone and Sorella, Mara and Tanasache, Florin D and Palleschi, Alessia and Ciccotelli, Claudio and Sakkalis, Vangelis and others},
  booktitle={2020 42nd Annual International Conference of the IEEE Engineering in Medicine \& Biology Society (EMBC)},
  pages={5705--5708},
  year={2020},
  organization={IEEE}
}

@inproceedings{pipyros2014cyber,
  title={A cyber attack evaluation methodology},
  author={Pipyros, Kosmas and Mitrou, Lilian and Gritzalis, Dimitris and Apostolopoulos, Theodore},
  booktitle={Proc. of the 13th European Conference on Cyber Warfare and Security},
  pages={264--270},
  year={2014}
}

@INPROCEEDINGS{9995441,
  author={Mahmud, M. Rasel and Stewart, Michael and Cordova, Alberto and Quarles, John},
  booktitle={2022 IEEE International Symposium on Mixed and Augmented Reality (ISMAR)}, 
  title={Auditory Feedback to Make Walking in Virtual Reality More Accessible}, 
  year={2022},
  volume={},
  number={},
  pages={847-856},
  doi={10.1109/ISMAR55827.2022.00103}}

@INPROCEEDINGS{9756779,
  author={Mahmud, M. Rasel and Stewart, Michael and Cordova, Alberto and Quarles, John},
  booktitle={2022 IEEE Conference on Virtual Reality and 3D User Interfaces (VR)}, 
  title={Auditory Feedback for Standing Balance Improvement in Virtual Reality}, 
  year={2022},
  volume={},
  number={},
  pages={782-791},
  doi={10.1109/VR51125.2022.00100}}

@article{mahmud2022vibrotactile,
  title={Vibrotactile feedback to make real walking in virtual reality more accessible},
  author={Mahmud, M Rasel and Stewart, Michael and Cordova, Alberto and Quarles, John},
  journal={arXiv preprint arXiv:2208.02403},
  year={2022}
}

@phdthesis{mahmud2023multimodal,
  title={Multimodal Feedback Techniques to Increase Accessibility of Immersive Virtual Reality},
  author={Mahmud, M Rasel},
  year={2023}
}

@inproceedings{mahmud2022standing,
  title={Standing balance improvement using vibrotactile feedback in virtual reality},
  author={Mahmud, M Rasel and Stewart, Michael and Cordova, Alberto and Quarles, John},
  booktitle={Proceedings of the 28th ACM Symposium on Virtual Reality Software and Technology},
  pages={1--11},
  year={2022}
}

@inproceedings{mahmud2023eyes,
  title={The Eyes Have It: Visual Feedback Methods to Make Walking in Immersive Virtual Reality More Accessible for People With Mobility Impairments While Utilizing Head-Mounted Displays},
  author={Mahmud, M Rasel and Cordova, Alberto and Quarles, John},
  booktitle={Proceedings of the 25th International ACM SIGACCESS Conference on Computers and Accessibility},
  pages={1--10},
  year={2023}
}

@article{mahmud2023visual,
  title={Visual cues for a steadier you: visual feedback methods improved standing balance in virtual reality for people with balance impairments},
  author={Mahmud, M Rasel and Cordova, Alberto and Quarles, John},
  journal={IEEE transactions on visualization and computer graphics},
  volume={29},
  number={11},
  pages={4666--4675},
  year={2023},
  publisher={IEEE}
}

@inproceedings{mahmud2023auditory,
  title={Auditory, Vibrotactile, or Visual? Investigating the Effective Feedback Modalities to Improve Standing Balance in Immersive Virtual Reality for People with Balance Impairments Due to Type 2 Diabetes},
  author={Mahmud, M Rasel and Cordova, Alberto and Quarles, John},
  booktitle={2023 IEEE International Symposium on Mixed and Augmented Reality (ISMAR)},
  pages={573--582},
  year={2023},
  organization={IEEE}
}

@article{mahmud2024multimodal,
  title={Multimodal feedback methods for advancing the accessibility of immersive virtual reality for people with balance impairments due to multiple sclerosis},
  author={Mahmud, M Rasel and Cordova, Alberto and Quarles, John},
  journal={IEEE Transactions on Visualization and Computer Graphics},
  year={2024},
  publisher={IEEE}
}

@inproceedings{cordova2023real,
  title={Real-time auditory feedback improves aging balance in immersive virtual environments},
  author={Cordova, Albeto and Stewart, Michael and Quarles, John and Mahmud, Rasel and Yao, Wan and Park, Se-Woong and Land, William and Ogu, David},
  booktitle={JOURNAL OF SPORT \& EXERCISE PSYCHOLOGY},
  volume={45},
  pages={S30--S30},
  year={2023},
  organization={HUMAN KINETICS PUBL INC 1607 N MARKET ST, PO BOX 5076, CHAMPAIGN, IL 61820~…}
}


\appendix

\end{document}